\documentclass[11pt, a4paper]{article}

\usepackage{jheppub, slashed, color}

\DeclareFontFamily{U}{MnSymbolC}{}
\DeclareSymbolFont{MnSyC}{U}{MnSymbolC}{m}{n}
\DeclareFontShape{U}{MnSymbolC}{m}{n}{
    <-6>  MnSymbolC5
   <6-7>  MnSymbolC6
   <7-8>  MnSymbolC7
   <8-9>  MnSymbolC8
   <9-10> MnSymbolC9
  <10-12> MnSymbolC10
  <12->   MnSymbolC12}{}
\DeclareMathSymbol{\intprod}{\mathbin}{MnSyC}{'270}

\newcommand{\del}{\partial}
\newcommand{\delb}{{\bar\partial}}
\newcommand{\Ds}{\slashed{D}}
\newcommand{\dels}{\slashed{\del}}

\newcommand{\vol}{\mathrm{vol}}

\newcommand{\Hom}{\mathop{\mathrm{Hom}}\nolimits}
\newcommand{\Map}{\mathop{\mathrm{Map}}\nolimits}

\renewcommand{\Im}{\mathop{\mathrm{Im}}\nolimits}
\renewcommand{\Re}{\mathop{\mathrm{Re}}\nolimits}

\newcommand{\Tr}{\mathop{\mathrm{Tr}}\nolimits}

\newcommand{\SU}{\mathrm{SU}}

\newcommand{\U}{\mathrm{U}}

\newcommand{\iso}{\cong}

\newcommand{\Z}{\mathbb{Z}}
\newcommand{\Q}{\mathbb{Q}}
\newcommand{\R}{\mathbb{R}}
\newcommand{\C}{\mathbb{C}}

\let\nc\newcommand
\let\renc\renewcommand
\nc{\wbar}{\overline}
\let\td\tilde
\let\wtd\widetilde
\let\wht\widehat
\let\mcl\mathcal

\nc{\ab}{{\bar{a}}} \nc{\at}{\tilde{a}} \nc{\ah}{\hat{a}}
\nc{\bb}{{\bar{b}}} \nc{\bt}{\tilde{b}} \nc{\bh}{\hat{b}}
\nc{\cb}{{\bar{c}}} \nc{\ct}{\tilde{c}} 
\nc{\db}{{\bar{d}}} \nc{\dt}{\tilde{d}} \renc{\dh}{\hat{d}}
\nc{\eb}{{\bar{e}}} \nc{\et}{\tilde{e}} \nc{\eh}{\hat{e}}
\nc{\fb}{{\bar{f}}} \nc{\ft}{\tilde{f}} \nc{\fh}{\hat{f}}
\nc{\gb}{{\bar{g}}} \nc{\gt}{\tilde{g}} \nc{\gh}{\hat{g}}
\nc{\hb}{{\bar{h}}} \nc{\hh}{\hat{h}} 
\nc{\ib}{{\bar{\imath}}} \nc{\ih}{\hat{\imath}} 
\nc{\jb}{{\bar{\jmath}}} \nc{\jt}{\tilde{\jmath}} \nc{\jh}{\hat{\jmath}}
\nc{\kb}{{\bar{k}}} \nc{\kt}{\tilde{k}} \nc{\kh}{\hat{k}}
\nc{\lb}{{\bar{l}}} \nc{\lt}{\tilde{l}} \nc{\lh}{\hat{l}}
\nc{\mb}{{\bar{m}}} \nc{\mt}{\tilde{m}} \nc{\mh}{\hat{m}}
\nc{\nb}{{\bar{n}}} \nc{\nt}{\tilde{n}} \nc{\nh}{\hat{n}}
\nc{\ob}{{\bar{o}}} \nc{\ot}{\tilde{o}} \nc{\oh}{\hat{o}}
\nc{\pb}{{\bar{p}}} \nc{\pt}{\tilde{p}} \nc{\ph}{\hat{p}}
\nc{\qb}{{\bar{q}}} \nc{\qt}{\tilde{q}} \nc{\qh}{\hat{q}}
\nc{\rb}{{\bar{r}}} \nc{\rt}{\tilde{r}} \nc{\rh}{\hat{r}}
\renc{\sb}{{\bar{s}}} \nc{\st}{\tilde{s}} \nc{\sh}{\hat{s}}
\nc{\tb}{{\bar{t}}} \renc{\th}{\hat{t}} 
\nc{\ub}{{\bar{u}}} \nc{\ut}{\tilde{u}} \nc{\uh}{\hat{u}}
\nc{\vb}{{\bar{v}}} \nc{\vt}{\tilde{v}} \nc{\vh}{\hat{v}}
\nc{\wb}{{\bar{w}}} \nc{\wt}{\tilde{w}} \nc{\wh}{\hat{w}}
\nc{\xb}{{\bar{x}}} \nc{\xt}{\tilde{x}} \nc{\xh}{\hat{x}}
\nc{\yb}{{\bar{y}}} \nc{\yt}{\tilde{y}} \nc{\yh}{\hat{y}}
\nc{\zb}{{\bar{z}}} \nc{\zt}{\tilde{z}} \nc{\zh}{\hat{z}}

\nc{\Ab}{\wbar{A}} \nc{\At}{\wtd{A}} \nc{\Ah}{\wht{A}}
\nc{\Bb}{\wbar{B}} \nc{\Bt}{\wtd{B}} \nc{\Bh}{\wht{B}}
\nc{\Cb}{\wbar{C}} \nc{\Ct}{\wtd{C}} \nc{\Ch}{\wht{C}}
\nc{\Db}{\wbar{D}} \nc{\Dt}{\wtd{D}} \nc{\Dh}{\wht{D}}
\nc{\Eb}{\wbar{E}} \nc{\Et}{\wtd{E}} \nc{\Eh}{\wht{E}}
\nc{\Fb}{\wbar{F}} \nc{\Ft}{\wtd{F}} \nc{\Fh}{\wht{F}}
\nc{\Gb}{\wbar{G}} \nc{\Gt}{\wtd{G}} \nc{\Gh}{\wht{G}}
\nc{\Hb}{\wbar{H}} \nc{\Ht}{\wtd{H}} \nc{\Hh}{\wht{H}}
\nc{\Ib}{\wbar{I}} \nc{\It}{\wtd{I}} \nc{\Ih}{\wht{I}}
\nc{\Jb}{\wbar{J}} \nc{\Jt}{\wtd{J}} \nc{\Jh}{\wht{J}}
\nc{\Kb}{\wbar{K}} \nc{\Kt}{\wtd{K}} \nc{\Kh}{\wht{K}}
\nc{\Lb}{\wbar{L}} \nc{\Lt}{\wtd{L}} \nc{\Lh}{\wht{L}}
\nc{\Mb}{\wbar{M}} \nc{\Mt}{\wtd{M}} \nc{\Mh}{\wht{M}}
\nc{\Nb}{\wbar{N}} \nc{\Nt}{\wtd{N}} \nc{\Nh}{\wht{N}}
\nc{\Ob}{\wbar{O}} \nc{\Ot}{\wtd{O}} \nc{\Oh}{\wht{O}}
\nc{\Pb}{\wbar{P}} \nc{\Pt}{\wtd{P}} \nc{\Ph}{\wht{P}}
\nc{\Qb}{\wbar{Q}} \nc{\Qt}{\wtd{Q}} \nc{\Qh}{\wht{Q}}
\nc{\Rb}{\wbar{R}} \nc{\Rt}{\wtd{R}} \nc{\Rh}{\wht{R}}
\nc{\Sb}{\wbar{S}} \nc{\St}{\wtd{S}} \nc{\Sh}{\wht{S}}
\nc{\Tb}{\wbar{T}} \nc{\Tt}{\wtd{T}} \nc{\Th}{\wht{T}}
\nc{\Ub}{\wbar{U}} \nc{\Ut}{\wtd{U}} \nc{\Uh}{\wht{U}}
\nc{\Vb}{\wbar{V}} \nc{\Vt}{\wtd{V}} \nc{\Vh}{\wht{V}}
\nc{\Wb}{\wbar{W}} \nc{\Wt}{\wtd{W}} \nc{\Wh}{\wht{W}}
\nc{\Xb}{\wbar{X}} \nc{\Xt}{\wtd{X}} \nc{\Xh}{\wht{X}}
\nc{\Yb}{\wbar{Y}} \nc{\Yt}{\wtd{Y}} \nc{\Yh}{\wht{Y}}
\nc{\Zb}{\wbar{Z}} \nc{\Zt}{\wtd{Z}} \nc{\Zh}{\wht{Z}}

\nc{\CA}{\mcl{A}} \nc{\CAb}{\wbar{\CA}} \nc{\CAt}{\wtd{\CA}} \nc{\CAh}{\wht{\CA}}
\nc{\CB}{\mcl{B}} \nc{\CBb}{\wbar{\CB}} \nc{\CBt}{\wtd{\CB}} \nc{\CBh}{\wht{\CB}}
\nc{\CC}{\mcl{C}} \nc{\CCb}{\wbar{\CC}} \nc{\CCt}{\wtd{\CC}} \nc{\CCh}{\wht{\CC}}
\nc{\cD}{\mcl{D}} \nc{\cDb}{\wbar{\cD}} \nc{\cDt}{\wtd{\cC}} \nc{\cDh}{\wht{\cD}}
\nc{\CE}{\mcl{E}} \nc{\CEb}{\wbar{\CE}} \nc{\CEt}{\wtd{\CE}} \nc{\CEh}{\wht{\CE}}
\nc{\CF}{\mcl{F}} \nc{\CFb}{\wbar{\CF}} \nc{\CFt}{\wtd{\CF}} \nc{\CFh}{\wht{\CF}}
\nc{\CG}{\mcl{G}} \nc{\CGb}{\wbar{\CG}} \nc{\CGt}{\wtd{\CG}} \nc{\CGh}{\wht{\CG}}
\nc{\CH}{\mcl{H}} \nc{\CHb}{\wbar{\CH}} \nc{\CHt}{\wtd{\CH}} \nc{\CHh}{\wht{\CH}}
\nc{\CI}{\mcl{I}} \nc{\CIb}{\wbar{\CI}} \nc{\CIt}{\wtd{\CI}} \nc{\CIh}{\wht{\CI}}
\nc{\CJ}{\mcl{J}} \nc{\CJb}{\wbar{\CJ}} \nc{\CJt}{\wtd{\CJ}} \nc{\CJh}{\wht{\CJ}}
\nc{\CK}{\mcl{K}} \nc{\CKb}{\wbar{\CK}} \nc{\CKt}{\wtd{\CK}} \nc{\CKh}{\wht{\CK}}
\nc{\CL}{\mcl{L}} \nc{\CLb}{\wbar{\CL}} \nc{\CLt}{\wtd{\CL}} \nc{\CLh}{\wht{\CL}}
\nc{\CM}{\mcl{M}} \nc{\CMb}{\wbar{\CM}} \nc{\CMt}{\wtd{\CM}} \nc{\CMh}{\wht{\CM}}
\nc{\CN}{\mcl{N}} \nc{\CNb}{\wbar{\CN}} \nc{\CNt}{\wtd{\CN}} \nc{\CNh}{\wht{\CN}}
\nc{\CO}{\mcl{O}} \nc{\COb}{\wbar{\CO}} \nc{\COt}{\wtd{\CO}} \nc{\COh}{\wht{\CO}}
\nc{\CQ}{\mcl{Q}} \nc{\CQb}{\wbar{\CQ}} \nc{\CQt}{\wtd{\CQ}} \nc{\CQh}{\wht{\CQ}}
\nc{\CR}{\mcl{R}} \nc{\CRb}{\wbar{\CR}} \nc{\CRt}{\wtd{\CR}} \nc{\CRh}{\wht{\CR}}
\nc{\CS}{\mcl{S}} \nc{\CSb}{\wbar{\CS}} \nc{\CSt}{\wtd{\CS}} \nc{\CSh}{\wht{\CS}}
\nc{\CT}{\mcl{T}} \nc{\CTb}{\wbar{\CT}} \nc{\CTt}{\wtd{\CT}} \nc{\CTh}{\wht{\CT}}
\nc{\CU}{\mcl{U}} \nc{\CUb}{\wbar{\CU}} \nc{\CUt}{\wtd{\CU}} \nc{\CUh}{\wht{\CU}}
\nc{\CV}{\mcl{V}} \nc{\CVb}{\wbar{\CV}} \nc{\CVt}{\wtd{\CV}} \nc{\CVh}{\wht{\CV}}
\nc{\CW}{\mcl{W}} \nc{\CWb}{\wbar{\CW}} \nc{\CWt}{\wtd{\CW}} \nc{\CWh}{\wht{\CW}}
\nc{\CX}{\mcl{X}} \nc{\CXb}{\wbar{\CX}} \nc{\CXt}{\wtd{\CX}} \nc{\CXh}{\wht{\CX}}
\nc{\CY}{\mcl{Y}} \nc{\CYb}{\wbar{\CY}} \nc{\CYt}{\wtd{\CY}} \nc{\CYh}{\wht{\CY}}
\nc{\CZ}{\mcl{Z}} \nc{\CZb}{\wbar{\CZ}} \nc{\CZt}{\wtd{\CZ}} \nc{\CZh}{\wht{\CZ}}

\let\eps\epsilon
\let\ups\upsilon
\let\veps\varepsilon
\let\vtht\vartheta
\let\vsgm\varsigma
\let\vphi\varphi
\let\vrho\varrho

\nc{\alphab}{\bar{\alpha}} \nc{\alphat}{\td{\alpha}} \nc{\alphah}{\hat{\alpha}}
\nc{\betab}{\bar{\beta}}   \nc{\betat}{\td{\beta}}   \nc{\betah}{\hat{\beta}} 
\nc{\gammab}{\bar{\gamma}} \nc{\gammat}{\td{\gamma}} \nc{\gammah}{\hat{\gamma}} 
\nc{\deltab}{\bar{\delta}} \nc{\deltat}{\td{\delta}} \nc{\deltah}{\hat{\delta}} 
\nc{\epsilonb}{\bar{\eps}} \nc{\epsilont}{\td{\eps}} \nc{\epsilonh}{\hat{\eps}} 
\nc{\vepsb}{\bar{\veps}}   \nc{\vepst}{\td{\veps}}   \nc{\vepsh}{\hat{\veps}} 
\nc{\zetab}{\bar{\zeta}}   \nc{\zetat}{\td{\zeta}}   \nc{\zetah}{\hat{\zeta}} 
\nc{\etab}{\bar{\eta}}     \nc{\etat}{\td{\eta}}     \nc{\etah}{\hat{\eta}} 
\nc{\thetab}{\bar{\theta}} \nc{\thetat}{\td{\theta}} \nc{\thetah}{\hat{\theta}} 
\nc{\vthetab}{\bar{\vtht}} \nc{\vthetat}{\td{\vtht}} \nc{\vthetah}{\hat{\vtht}} 
\nc{\lambdab}{\bar{\lambda}} \nc{\lambdat}{\td{\lambda}} \nc{\lambdah}{\hat{\lambda}} 
\nc{\iotab}{\bar{\iota}}   \nc{\iotat}{\td{\iota}}   \nc{\iotah}{\hat{\iota}} 
\nc{\kappab}{\bar{\kappa}} \nc{\kappat}{\td{\kappa}} \nc{\kappah}{\hat{\kappa}} 
\nc{\lmdb}{\bar{\lmd}}     \nc{\lmdt}{\td{\lmd}}     \nc{\lmdh}{\hat{\lmd}} 
\nc{\mub}{\bar{\mu}}       \nc{\mut}{\td{\mu}}       \nc{\muh}{\hat{\mu}} 
\nc{\nub}{\bar{\nu}}       \nc{\nut}{\td{\nu}}       \nc{\nuh}{\hat{\nu}} 
\nc{\xib}{\bar{\xi}}       \nc{\xit}{\td{\xi}}       \nc{\xih}{\hat{\xi}} 
\nc{\pib}{\bar{\pi}}       \nc{\pit}{\td{\pi}}       \nc{\pih}{\hat{\pi}} 
\nc{\vpib}{\bar{\vpi}}     \nc{\vpit}{\td{\vpi}}     \nc{\vpih}{\hat{\vpi}} 
\nc{\rhob}{\bar{\rho}}     \nc{\rhot}{\td{\rho}}     \nc{\rhoh}{\hat{\rho}} 
\nc{\vrhob}{\bar{\vrho}}   \nc{\vrhot}{\td{\vrho}}   \nc{\vrhoh}{\hat{\vrho}} 
\nc{\sigmab}{\bar{\sigma}} \nc{\sigmat}{\td{\sigma}} \nc{\sigmah}{\hat{\sigma}} 
\nc{\vsigmab}{\bar{\vsgm}} \nc{\vsigmat}{\td{\vsgm}} \nc{\vsigmah}{\hat{\vsgm}} 
\nc{\taub}{\bar{\tau}}     \nc{\taut}{\td{\tau}}     \nc{\tauh}{\hat{\tau}} 
\nc{\upsb}{\bar{\ups}} \nc{\upst}{\td{\ups}} \nc{\upsh}{\hat{\ups}} 
\nc{\phib}{\bar{\phi}}     \nc{\phit}{\td{\phi}}     \nc{\phih}{\hat{\phi}} 
\nc{\varphib}{\bar{\vphi}}   \nc{\varphit}{\td{\vphi}}   \nc{\varphih}{\hat{\vphi}} 
\nc{\chib}{\bar{\chi}}     \nc{\chit}{\td{\chi}}     \nc{\chih}{\hat{\chi}} 
\nc{\psib}{\bar{\psi}}     \nc{\psit}{\wtd{\psi}}     \nc{\psih}{\hat{\psi}} 
\nc{\omegab}{\bar{\omega}} \nc{\omegat}{\td{\omega}} \nc{\omegah}{\hat{\omega}} 

\nc{\Gammab}{\wbar{\Gamma}}     \nc{\Gammat}{\wtd{\Gamma}}     \nc{\Gammah}{\wht{\Gamma}}
\nc{\Deltab}{\wbar{\Delta}}     \nc{\Deltat}{\wtd{\Delta}}     \nc{\Deltah}{\wht{\Delta}}
\nc{\Thetab}{\wbar{\Theta}}     \nc{\Thetat}{\wtd{\Theta}}     \nc{\Thetah}{\wht{\Theta}}
\nc{\Lambdab}{\wbar{\Lambda}}   \nc{\Lambdat}{\wtd{\Lambda}}   \nc{\Lambdah}{\wht{\Lambda}}
\nc{\Xib}{\wbar{\Xi}}           \nc{\Xit}{\wtd{\Xi}}           \nc{\Xih}{\wht{\Xi}}
\nc{\Pib}{\wbar{\Pi}}           \nc{\Pit}{\wtd{\Pi}}           \nc{\Pih}{\wht{\Pi}}
\nc{\Sigmab}{\wbar{\Sigma}}     \nc{\Sigmat}{\wtd{\Sigma}}     \nc{\Sigmah}{\wht{\Sigma}}
\nc{\Upsilonb}{\wbar{\Upsilon}} \nc{\Upsilont}{\wtd{\Upsilon}} \nc{\Upsilonh}{\wht{\Upsilon}}
\nc{\Phib}{\wbar{\Phi}}         \nc{\Phit}{\wtd{\Phi}}         \nc{\Phih}{\wht{\Phi}}
\nc{\Psib}{\wbar{\Psi}}         \nc{\Psit}{\wtd{\Psi}}         \nc{\Psih}{\wht{\Psi}}
\nc{\Omegab}{\wbar{\Omega}}     \nc{\Omegat}{\wtd{\Omega}}     \nc{\Omegah}{\wht{\Omega}}

\newcommand{\rmd}{\mathrm{d}}
\newcommand{\nablas}{\slashed{\nabla}}

\title{$\CN = 2$ supersymmetric gauge theories and quantum integrable
  systems}

\author[a]{Yuan Luo,}
\author[a]{Meng-Chwan Tan}
\author[a, b,c]{and Junya Yagi}

\emailAdd{yuanluo@nus.edu.sg, mctan@nus.edu.sg, junya.yagi@sissa.it}
\affiliation[a]{Department of Physics, National University of
  Singapore \\  2 Science Drive 3, Singapore 117551}
\affiliation[b]{International School for Advanced Studies (SISSA) \\
Via Bonomea, 265, 34136 Trieste, Italy}
\affiliation[c]{INFN, Sezione di Trieste \\
Via Valerio, 2, 34149 Trieste, Italy}

\abstract{We study $\CN = 2$ supersymmetric gauge theories on the
  product of a two-sphere and a cylinder.  We show that the low-energy
  dynamics of a BPS sector of such a theory is described by a quantum
  integrable system, with the Planck constant set by the inverse of
  the radius of the sphere.  If the sphere is replaced with a
  hemisphere, then our system reduces to an integrable system of the
  type studied by Nekrasov and Shatashvili.  In this case we establish
  a correspondence between the effective prepotential of the gauge
  theory and the Yang-Yang function of the integrable system.}


\arxivnumber{}

\renewcommand{\ell}{l}

\begin{document}
\maketitle
\flushbottom

\section{Introduction}

It was not long after the seminal work of Seiberg and Witten
\cite{Seiberg:1994rs, Seiberg:1994aj} when people realized a
connection between $\CN = 2$ supersymmetric gauge theories in four
dimensions and complex integrable systems \cite{Gorsky:1995zq,
  Martinec:1995by, Nakatsu:1995bz, Donagi:1995cf, Martinec:1995qn,
  Gorsky:1995sr, Itoyama:1995nv, Itoyama:1995uj}.  A few years ago,
Nekrasov and Shatashvili \cite{Nekrasov:2009rc} found that turning on
$\Omega$-deformation \cite{Nekrasov:2002qd} on a two-plane quantizes
these integrable systems, with the deformation parameter $\veps$
playing the role of the Planck constant.  An explanation of this
result was subsequently given by Nekrasov and Witten
\cite{Nekrasov:2010ka} using a brane construction.

In this paper we establish another, yet closely related, connection
between $\CN = 2$ supersymmetric gauge theories and quantum integrable
systems.  Instead of turning on $\Omega$-deformation, we compactify a
two-plane to a round two-sphere $S^2$ of radius $r$.  One of the
remaining two dimensions is compactified to a circle $S^1$ of radius
$R$; therefore our setup is an $\CN = 2$ supersymmetric gauge theory
formulated on $S^2 \times \R \times S^1$.  We will show that the
low-energy dynamics of a BPS sector of this theory is described by a
quantum integrable system, with the Planck constant set by $1/r$.
This system quantizes the real integrable system whose symplectic form
is $\Re\Omega$, where $\Omega$ is the holomorphic symplectic form of
the complex integrable system associated to the Coulomb branch.

The logic of our argument is simple.  First, we go to the effective
three-dimensional description at energies $\mu \ll 1/R$.  After
dualization of the gauge fields to periodic scalars, we get an $\CN =
4$ supersymmetric sigma model on $S^2 \times \R$, whose target space
$\CM$ is the total space of the complex integrable system
\cite{Seiberg:1996nz, Gaiotto:2008cd}.  Then, we localize the path
integral for the BPS sector of this sigma model and reduce it to the
path integral for the quantum integrable system.

Our system is in a sense twice as big as Nekrasov and Shatashvili's:
theirs is essentially the restriction of ours to a middle-dimensional
submanifold that is Lagrangian with respect to $\Im\Omega$.  If we
replace the $S^2$ with a hemisphere, then we obtain the restricted
system for a suitable boundary condition.  We will also discuss this
construction, and establish a correspondence between the effective
prepotential of the gauge theory and the Yang-Yang function of the
restricted system.

The rest of the paper is organized as follows.  Section \ref{SW-CIS}
is a review of background materials.  In section \ref{QIS} we present
our derivation of the connection between the BPS sector of the
low-energy effective theory and the quantum integrable system.  We
consider the hemisphere case in section~\ref{hemisphere}.  The
construction of the ultraviolet theory is treated in appendix
\ref{N=2-SGT}, where we formulate $\CN = 2$ supersymmetric gauge
theories more generally on the product of $S^2$ and any Riemann
surface.

\section{Seiberg-Witten theory and complex integrable systems}
\label{SW-CIS}

To begin, let us review the basic elements that enter our story.  We
consider an $\CN = 2$ supersymmetric gauge theory on flat spacetime
$\R^4$, with gauge group of rank $r$ and a characteristic mass scale
$\Lambda$.  After recalling the structure of the low-energy effective
theory, we explain how it is encoded in a complex integrable system,
and how this system emerges as the target space of the sigma model
obtained by compactification on $S^1$.  To keep the discussion simple,
we will ignore flavor symmetries for the most part.  Their effects are
briefly discussed at the end of the section.

\subsection{Seiberg-Witten theory}

We are interested in the effective description of the theory on the
Coulomb branch at energies $\mu \ll \Lambda$.  The Coulomb branch is
parametrized by the vacuum expectation values of the gauge-invariant
polynomials in the vector multiplet scalar $\phi$.  There are $r$ such
parameters, providing coordinates for a complex manifold $\CB$.

At each point $u \in \CB$, the gauge group is broken to a maximal
torus $\U(1)^r$, and there is a lattice $\Gamma_u \subset \R^{2r}$ of
electric and magnetic charges.  The lattice is equipped with a
nondegenerate skew-symmetric bilinear form
\begin{equation}
  \langle \ , \ \rangle\colon \Gamma_u \times \Gamma_u \to \Z,
\end{equation}
which is $\Z$-valued by the Dirac quantization condition.  The charge
lattices at the different points of $\CB$ form a fibration
\begin{equation}
  \Gamma \to \CB.
\end{equation}
The fibration has nontrivial monodromy around the singular loci in
$\CB$ of complex codimension $1$.

Locally on $\CB$, one can find a symplectic basis $\{\alpha_I,
\beta^I\} \subset \Gamma$, $I = 1$, $\dotsc$, $r$, which satisfy
\begin{equation}
  \langle \alpha_I, \alpha_J\rangle
  = \langle \beta^I, \beta^J\rangle = 0, \quad
  \langle \alpha_I, \beta^J\rangle = d_I \delta_I^J,
\end{equation}
with $d_I$ positive integers such that $d_I$ divides $d_{I+1}$.  Such
a choice determines a duality frame, that is, a local splitting of
$\Gamma$ into the Lagrangian sublattices $\Gamma_m$ and $\Gamma_e$ of
magnetic and electric charges, generated by $\{\alpha_I\}$ and
$\{\beta^I\}$, respectively.

We denote by $\Gamma^*$ the fibration over $\CB$ whose fiber at $u \in
\CB$ is the dual lattice $\Gamma_u^*$ of $\Gamma_u$, which is the
lattice consisting of $x \in \R^{2r}$ such that $\langle \gamma,
x\rangle \in \Z$ for all $\gamma \in \Gamma$.  We have $\Gamma \subset
\Gamma^*$, and there is a natural $\Q$-valued pairing on $\Gamma^*$
which extends the pairing on $\Gamma$.  The homomorphism $x \mapsto
\langle \ , x\rangle$ gives an isomorphism
\begin{equation}
  \Gamma_u^* \iso \Hom(\Gamma_u, \Z).
\end{equation}
Concretely, if we set $\alpha^I = \beta^I/d_I$, \ $\beta_I =
-\alpha_I/d_I$, then these generate $\Gamma_u^*$ and are mapped to the
dual basis of $\Hom(\Gamma_u, \Z)$.

For simplicity we will assume that $(d_1, \dotsc, d_r) = (1, \dotsc,
1)$, in other words, all charges allowed by the Dirac quantization
condition actually appear in the theory.  Then the dual basis
$\{\alpha^I, \beta_I\}$ is given by
\begin{equation}
  \alpha^I = \beta^I, \qquad
  \beta_I = -\alpha_I,
\end{equation}
and we have
\begin{equation}
  \Gamma^* = \Gamma.
\end{equation}
The generalization to the case of $(d_1, \dotsc, d_r) \neq (1, \dotsc,
1)$ is not hard.

The mass of a particle of charge $\gamma \in \Gamma_u$ is bounded from
below by the absolute value of the central charge $Z_\gamma(u)$, which
is a holomorphic function on $\CB$ satisfying $Z_{\gamma_1 +
  \gamma_2}(u) = Z_{\gamma_1}(u) + Z_{\gamma_2}(u)$.  Letting $\gamma$
and $u$ vary, we get a homomorphism
\begin{equation}
  Z\colon \Gamma \to \C.
\end{equation}
It satisfies the nondegeneracy condition
\begin{equation}
  \label{ND}
  \langle \rmd Z, \rmd\Zb\rangle > 0
\end{equation}
and the transversality condition
\begin{equation}
  \label{TC}
  \langle \rmd Z, \rmd Z\rangle = 0,
\end{equation}
where the wedge product of differential forms is implicit.

To understand the meaning of these conditions, let us locally choose a
symplectic basis and write
\begin{equation}
  Z = a^I\beta_I + a_{D,I} \alpha^I
\end{equation}
with some locally-defined holomorphic functions $a^I$, $a_{D,I}$ on
$\CB$.  Then the nondegeneracy condition \eqref{ND} reads
\begin{equation}
  \Re\bigl(\rmd a^I \wedge \rmd\ab_{D,I}\bigr)  < 0.
\end{equation}
In particular, this implies that the matrices $(\del a^I/\del u^J)$
and $(\del a_{D,I}/\del u^J)$ are invertible for any holomorphic
coordinates $u^I$ on $\CB$.  Thus the $a^I$ give local holomorphic
coordinates on $\CB$, and so do the $a_{D,I}$.  These are called
special coordinates.  On the other hand, the transversality condition
\eqref{TC} can be written as
\begin{equation}
  \rmd\bigl(a_{D,I} \rmd a^I\bigr) = 0.
\end{equation}
This ensures that locally there is a holomorphic function $\CF$ such
that $a_{D,I} \rmd a^I = \rmd\CF$.  The prepotential $\CF$ relates the
special coordinates $a^I$ and $a_{D,I}$ by
\begin{equation}
  a_{D,I} = \frac{\del\CF}{\del a^I}.
\end{equation}

We interpret the positive $(1,1)$-form $-\Re(\rmd a^I \wedge
\rmd\ab_{D,I})$ as a K\"ahler form on $\CB$.  If we define the period
matrix $\tau = (\tau_{IJ})$ by
\begin{equation}
  \tau_{IJ} = \frac{\del a_{D,I}}{\del a^J}
           = \frac{\del^2\CF}{\del a^I \del a^J},
\end{equation}
then by the nondegeneracy condition
\begin{equation}
  \Im \tau > 0.
\end{equation}
Finally, the bosonic part of the effective Lagrangian is given by
\begin{equation}
  \label{L}
  \CL = \frac{1}{4\pi} \Im\tau_{IJ}\bigl(
        \rmd a^I \wedge \star \rmd \ab^J + F^I \wedge \star F^J\bigr)
        + \frac{i}{4\pi} \Re\tau_{IJ} F^I \wedge F^J.
\end{equation}
In this expression, $a^I$ are vector multiplet scalars whose vacuum
expectation values at $u \in \CB$ give the special coordinates
$a^I(u)$, and $F^I = \rmd A^I$ are the gauge field strengths.

\subsection{Seiberg-Witten integrable system}

The structure of the Coulomb branch naturally leads to a complex
integrable system \cite{Donagi:1995cf}.  To establish this connection
we consider the fibration
\begin{equation}
  \CMt = \Gamma \otimes_\Z \R/\Z \to \CB,
\end{equation}
whose fibers are $2r$-tori.

Choosing a local symplectic basis $\{\alpha_I, \beta^I\}$ of $\Gamma$,
we write a point $\vartheta$ in the fiber $\CMt_u$ as
\begin{equation}
  \vartheta = \vartheta_m^I \alpha_I + \vartheta_{e,I} \beta^I.
\end{equation}
Then $(\vartheta_m^I, \vartheta_{e,I})$ are periodic coordinates on
$\CMt_u$.  There is an isomorphism $H_1(\CMt_u; \Z) \to \Gamma_u$
given by
\begin{equation}
  \label{H1Gamma}
  \gamma \mapsto \oint_\gamma \rmd\vartheta.
\end{equation}
Under this isomorphism, the duals $\alpha^I$, $\beta_I \in \Gamma_u^*$
of $\alpha_I$, $\beta^I$ are identified with classes in $H^1(\CMt_u;
\Z)$ represented by the one-forms $\rmd\vartheta_m^I$,
$\rmd\vartheta_{e,I}$.  We introduce complex coordinates $w_I$ on
$\CMt_u$ by
\begin{equation}
  w_I = \vartheta_{e,I} + \tau_{IJ} \vartheta_m^J,
\end{equation}
so that the pairing $\langle \ , \ \rangle$ is represented by the
negative $(1,1)$-form
\begin{equation}
  -\frac{i}{2} (\Im\tau)^{-1,IJ} \rmd w_I \wedge \rmd\wb_J
  = \rmd\vartheta_m^I \wedge \rmd\vartheta_{e,I}.
\end{equation}
This turns $\CMt_u$ into a (principally polarized) abelian variety, which
is a complex torus that can be described by algebraic equations.  

The central charge $Z(u)\colon \Gamma_u \to \C$ is pulled back by the
isomorphism \eqref{H1Gamma} to the class in $H^1(\CMt_u; \C)$
represented by $Z(u) \cdot \rmd\vartheta = a^I \rmd\vartheta_{e,I} +
a_{D,I} \rmd\vartheta_m^I$.  The derivative of the one-form $Z \cdot
\rmd\vartheta$ on the total space $\CMt$ is the holomorphic two-form
\begin{equation}
  \Omega = \rmd Z \cdot \rmd\vartheta = \rmd a^I \wedge \rmd w_I.
\end{equation}
Here we used the relation $\del\tau_{IJ}/\del a^K = \del\tau_{KJ}/\del
a^I$.

Since $\Omega$ is closed and nondegenerate, it is a holomorphic
symplectic form on $\CMt$.  The fibers of $\CMt$ are Lagrangian
subvarieties with respect to $\Omega$.  The associated Poisson
brackets are
\begin{equation}
  \{a^I, a^J\} = \{w_I, w_J\} = 0, \qquad
  \{a^I, w_J\} = \delta^I_J.
\end{equation}
There are $r$ independent Poisson-commuting complex quantities $a^I$
in the phase space $\CMt$ of complex dimension $2r$.  Hence, the
fibration $\CMt \to \CB$ describes an integrable system in the complex
sense.

\subsection{Compactification to three dimensions}

The complex integrable system described above is not merely a fancy
way of encoding the low-energy physics.  Actually, it emerges as the
target space when the theory is compactified on a circle
\cite{Seiberg:1996nz}.

We compactify the $x^4$-direction to a circle $S^1$ of radius $R$.  We
take $R \gg 1/\Lambda$.  Then, the dynamics at low energies $\mu \ll
\Lambda$ but still $\mu \gg 1/R$ is described by essentially the same
effective theory as we considered previously, formulated this time on
$\R^3 \times S^1$ rather than $\R^4$, and possibly with finite-size
corrections to $\CF$ which vanish in the limit $R \to \infty$.
Further in the infrared, at energies $\mu \ll 1/R$, the theory is
effectively three-dimensional.

This three-dimensional theory is not the simple dimensional reduction
of the effective theory on $\R^3 \times S^1$, even though the
Kaluza-Klein modes are very massive and decouple.  This is because
the latter theory supports topologically nontrivial configurations in
which the worldlines of BPS particles wrap the $S^1$.  Such
configurations appear as instantons in three dimensions.  The action
for these instantons is roughly $2\pi R|Z|$, and is not necessarily
large.

If $R|Z|$ is very large, however, the instanton effects are
suppressed.  Thus, for sufficiently large $R$, the effective
three-dimensional Lagrangian is obtained to leading order by
dimensional reduction of the four-dimensional Lagrangian, as far as
one stays away from the singular loci in $\CB$ where some BPS
particles become massless.  Let us look at this case and identify the
three-dimensional theory.

Dimensional reduction for the scalars $a^I$ is straightforward.  For
the gauge field, we note that at each point on the $\R^3$, the
components $A_4^I$ describe connections on line bundles over the
$S^1$.  Since connections on $S^1$ are determined up to gauge
transformations by their holonomies,
\begin{equation}
  \exp\biggl(i \oint A^I_4 \, \rmd x^4\biggr),
\end{equation}
we can account for the gauge freedom in the $x^4$-direction by setting
\begin{equation}
  \label{A3}
  A_4^I = \frac{\theta_e^I}{2\pi R},
\end{equation}
with $\theta_e^I$ periodic scalars with periodicity $2\pi$ that are
independent of $x^4$.  The residual gauge symmetry is given by the
gauge transformations on the $\R^3$.  Plugging the expression
\eqref{A3} into the effective Lagrangian \eqref{L}, dropping all the
$x^4$-dependence and integrating over the $x^4$-direction, we get the
three-dimensional Lagrangian
\begin{equation}
  \label{L3}
  \CL^{(3)} =
  \frac{R}{2} \Im\tau_{IJ}
  \biggl(\rmd a^I \wedge \star \rmd \ab^J + F^{(3),I} \wedge \star F^{(3),J}
  + \frac{\rmd\theta_e^I \wedge \star\rmd\theta_e^J}{4\pi^2 R^2}\biggr)
  + \frac{i}{2\pi} \Re\tau_{IJ} F^{(3),I} \wedge \rmd\theta_e^J.
\end{equation}
Here $F^{(3),I}$ are the field strengths of the gauge fields
$A^{(3),I}$, coming from the remaining components of $A^I$.

In three dimensions we can dualize gauge fields to scalars.  To do
this we convert the path integral variables from $A^{(3),I}$ to
$F^{(3),I}$.  The constraint $F^{(3),I}$ must obey is that through any
closed surface $S \subset \R^3$, their magnetic fluxes must be
integers:
\begin{equation}
  \label{FQ}
  \frac{1}{2\pi} \int_S F^{(3),I} \in \Z.
\end{equation}
(If $A^{(3),I}$ are connections on line bundles $L_I$, then
$F^{(3),I}/2\pi$ represent the first Chern classes $c_1(L_I) \in
H^1(S; \Z)$.)  So we introduce periodic scalars $\theta_{m,I}$ of
periodicity $2\pi$ as Lagrange multipliers, and add to the action the
term
\begin{equation}
  -\frac{i}{2\pi} \int_{\R^3} F^{(3),I} \wedge \rmd\theta_{m,I}.
\end{equation}
To see that integrating $\theta_{m,I}$ out produces the constraint
\eqref{FQ}, consider a continuous configuration such that
$\theta_{m,I}$ jump by $2\pi n_I$ for some $n_I \in \Z$ as we cross
$S$ from inside.  Then $\rmd\theta_{m,I}$ contain $2\pi n_I
\delta(S)$, where $\delta(S)$ is a two-form with delta-function
support on $S$ which represents the Poincar\'e dual of the homology
class $[S]$.  Thus the added term contains the factor
\begin{equation}
  -in_I \int_{S} F^{(3),I},
\end{equation}
and a summation over $n_I$ produces the desired constraint.

Integrating out $F^{(3),I}$ instead of $\theta_{m,I}$, we get the dualized
Lagrangian
\begin{equation}
  \label{LD3}
  \CL_D^{(3)} =
  \frac{R}{2} \Im\tau_{IJ} \bigl(
  \rmd a^I \wedge \star \rmd \ab^J
  + \eta^I \wedge \star\etab^J\bigr),
\end{equation}
with
\begin{equation}
  \eta^I = \frac{1}{2\pi R} (\Im\tau)^{-1,IJ}
             \bigl(\rmd\theta_{m,J} - \tau_{JK} \, \rmd\theta_e^K\bigr).
\end{equation}
This is the bosonic Lagrangian for a sigma model with target space
metric
\begin{equation}
  \label{SFM}
  g^{\text{sf}}
  = R \Im\tau_{IJ} \,
    \bigl(\rmd a^I \, \rmd \ab^J + \eta^I \, \etab^J\bigr).
\end{equation}
This ``semiflat'' metric $g^{\text{sf}}$ is singular over the singular
loci in $\CB$, around which $a^I$ have monodromies.  Instantons
correct $g^{\text{sf}}$ to a smooth metric $g$.

The theory has $\CN = 4$ supersymmetry in three dimensions, requiring
the target space $\CM$ of the sigma model to be a hyper\-k\"ahler
manifold.  This means that $\CM$ has three independent complex
structures $J_\alpha$, $\alpha = 1$, $2$, $3$, obeying the relation
\begin{equation}
  J^2_\alpha = J_1 J_2 J_3 = -1,
\end{equation}
and the metric $g$ is K\"ahler with respect to each $J_\alpha$.  In
the semiflat approximation, we can take $J_\alpha$ to act on $T^*\CM$
as follows:
\begin{equation}
  \begin{aligned}
    J_1&\colon (\rmd a^I, \eta^I)
               \mapsto (i\etab^I, -i\rmd\ab^I),
    \\
    J_2&\colon (\rmd a^I, \eta^I)
               \mapsto (-\etab^I, \rmd\ab^I),
    \\
    J_3&\colon (\rmd a^I, \eta^I) \mapsto (i\rmd a^I, i\eta^I).
  \end{aligned}
\end{equation}
One can check that the semiflat metric \eqref{SFM} is indeed K\"ahler
with respect to each of these complex structures.  Identifying the
exact hyperk\"ahler structure of $\CM$ is a difficult problem, and is
closely related to the wall-crossing phenomenon of BPS spectrum
\cite{Gaiotto:2008cd}.

So far we have described $\CM$ in some neighborhood of $\CB$ with a
chosen symplectic basis.  Globally, $\CM$ is a fibration over $\CB$
whose fibers are $2r$-tori parametrized by the periodic scalars
$(\theta_e^I, \theta_{m,I})$.  To better understand its geometry we
should go back to the four-dimensional description.  In four
dimensions we have the formula
\begin{equation}
  \label{theta_e}
  \theta_e^I = \oint_C A^I,
\end{equation}
where $C$ is a cycle located at a point in $\R^3$ and wrapped on
the $S^1$.  Choosing any surface $D$ such that $\del D = C$, we can
rewrite $\theta_e^I$ as the integration of $F^I$ over $D$.  On the
other hand, the dualization procedure in three dimensions sets
\begin{equation}
  \label{dtheta_m}
  \rmd\theta_{m,I}
  = \Re\tau_{IJ} \, \rmd\theta_e^J - 2\pi i R \Im\tau_{IJ} \star F^{(3),J}.
\end{equation}
This relation would follow if we define $\theta_{m,I}$ to be the
integral over $D$ of
\begin{equation}
  \label{F_D}
  F_{D,I} = \Re\tau_{IJ} \, F^J - i \Im\tau_{IJ} \star F^J.
\end{equation}
The equations of motion imply $\rmd F_{D,I} = 0$, so we can write
\begin{equation}
  \label{theta_m}
  \theta_{m,I} = \oint_C A_{D,I},
\end{equation}
using gauge fields $A_{D,I}$ for $F_{D,I}$.

As is clear from the symmetry between the equations $\rmd F^I = 0$ and
$\rmd F_{D,I} = 0$, the field strengths $F^I$ and $F_{D,I}$ are dual
to each other, and together form a $\Gamma^*$-valued two-form
$\mathbb{F} = F^I \beta_I + F_{D,I} \alpha^I$.
Similarly, the gauge fields $A^I$ and $A_{D,I}$ form a
$\Gamma^*$-valued gauge field $\mathbb{A} = A^I \beta_I + A_{D,I}
\alpha^I$.  So writing
\begin{equation}
  \theta = \theta_e^I \beta_I + \theta_{m,I} \alpha^I,
\end{equation}
we can combine the two formulas \eqref{theta_e} and \eqref{theta_m} into a
single formula that is independent of the choice of symplectic basis:
\begin{equation}
  \theta = \oint_C \mathbb{A}.
\end{equation}
Thus $\theta$ is a map to $\Gamma_a^* \otimes_\Z \R/2\pi\Z$, while the
$a^I$ give a map $a\colon \R^3 \to \CB$.

This consideration suggests $\CM \iso \Gamma^* \otimes_\Z \R/2\pi\Z$.
In turn, this space is isomorphic to the Seiberg-Witten fibration
$\CMt = \Gamma \otimes_\Z \R/\Z$ since $\Gamma^* = \Gamma$ by
assumption:
\begin{equation}
  \CM \iso \CMt.
\end{equation}
If we identify $\theta = 2\pi\vartheta$ under this isomorphism, then
we have the relations
\begin{equation}
  \theta_e^I = -2\pi\vartheta_m^I, \qquad
  \theta_{m,I} = 2\pi\vartheta_{e,I}.
\end{equation}
The holomorphic symplectic form $\Omega$ is identified as
\begin{equation}
  \label{Omega}
  \Omega = \frac{1}{2\pi} \rmd a^I \wedge \rmd z_I = -i(\omega_1 + i\omega_2),
\end{equation}
where we equipped the fibers with complex coordinates
\begin{equation}
  \label{z_I}
  z_I = \theta_{m,I} - \tau_{IJ} \theta_e^J = 2\pi w_I.
\end{equation}

In fact, it is not entirely true that $\CM$ is isomorphic to $\Gamma^*
\otimes_\Z \R/2\pi\Z$.  The reason is that whereas $\theta_e^I$ are
determined by the formula \eqref{theta_e}, the relation
\eqref{dtheta_m} determines the corresponding formula \eqref{theta_m}
only up to a constant.  Thus we have a collection of constants, each
associated to an open patch in $\CB$ equipped with a chosen symplectic
basis.  Locally we can discard these constants since the Lagrangian
depends on $\theta_{m,I}$ only through their derivatives.  Globally,
setting all of them to zero consistently may not be possible.  Indeed,
it was observed in \cite{Gaiotto:2008cd} that $\theta_{m,I}$ can have
monodromy shifting them by $\pi$.  Such monodromy does not affect the
fact that the fibration $\CM \to \CB$ defines an integrable system, as
it leaves the holomorphic symplectic form invariant.

What happens to the integrable system structure when the instanton
corrections are included?  The structure is associated with the
complex structure $J_3$.  It is special among all the complex
structures of $\CM$ in the sense that it is the only complex structure
under which $Z$ is holomorphic.  Instanton corrections are accompanied
with a factor of $\exp(-2\pi R|Z|)$, so cannot arise in quantities
that are holomorphic in $J_3$.  This implies that $J_3$ itself and the
associated holomorphic two-form $\Omega$, and hence also the
integrable system structure, are protected against the instanton
corrections.

\subsection{Flavor symmetries}

Let us briefly discuss what changes have to be made when the theory
has flavor symmetries.  For more discussions we refer the reader to
\cite{Gaiotto:2009hg, Donagi:1995cf, Donagi:1997sr}.

In the presence of flavor symmetries, the charge lattice $\Gamma$ is
equipped with a degenerate skew-symmetric bilinear form $\langle \ , \
\rangle$ whose radical is the lattice $\Gamma_f$ of flavor charges.
The quotient $\Gamma_g = \Gamma/\Gamma_f$ is the lattice of gauge
charges, on which $\langle \ , \ \rangle$ induces a symplectic
pairing.  The central charge homomorphism $Z\colon \Gamma \to \C$
varies holomorphically on $\CB$, and moreover $Z_\gamma$ is constant
for any $\gamma \in \Gamma_f$.  Thus $\rmd Z$ descends to a one-form
with values in $\Gamma_g^*$.  This is subject to the conditions
\eqref{ND} and $\eqref{TC}$.

Locally on $\CB$, we can decompose $\Gamma$ as $\Gamma = \Gamma'
\oplus \Gamma_f$, and choose a symplectic basis $\{\alpha_I, \beta^I\}$
of $\Gamma'$ and a basis $\{\gamma^i\}$ of $\Gamma_f$.  Then the
central charge can be written as
\begin{equation}
  Z = a^I \alpha_I + a_{D,I} \beta^I + m_i \gamma^i.
\end{equation}
The complex parameters $m_i$ are identified with the hypermultiplet
masses.  Monodromy around the singular loci in $\CB$ can shift the
duality frame by flavor charges, thereby shifting $a^I$, $a_{D,I}$ by
integral linear combinations of $m_i$.

In the framework of the Seiberg-Witten fibration $\CMt \to \CB$, the
presence of flavor symmetries removes codimension-$1$ subvarieties
$D_{i,u}$ from the fibers $\CMt_u$.  Letting $u$ vary these define
codimension-$1$ subvarieties $D_i$ in the total space.  The gauge
charges $\alpha_I$, $\beta^I$ are represented by cycles of $\CMt_u$
avoiding the $D_{i,u}$, and $\gamma^i$ are represented by cycles
encircling $D_{i,u}$.  The central charge $Z$ is now represented by a
one-form that contains terms meromorphic in $w_I$ with residues
$m_i/2\pi i$.  Its derivative thus contains delta functions.
Correspondingly, the holomorphic symplectic form $\Omega$ no longer
vanishes in the cohomology:
\begin{equation}
  [\Omega] = \sum_i m_i [D_i],
\end{equation}

\section{Quantum integrable systems from theories on $S^2 \times \R
  \times S^1$}
\label{QIS}

Now we replace two flat directions by a round two-sphere $S^2$ of
radius $r$, and study the low-energy effective theory on the geometry
$S^2 \times \R \times S^1$.  By localization of the path integral, we
will establish that a BPS sector of the effective theory is described
by a quantum integrable system.

\subsection{Ultraviolet theory}
\label{UV}

Our first task is to formulate $\CN = 2$ supersymmetric gauge theories
on $S^2 \times \R \times S^1$.  To this end we will treat a slightly
more general setup, in which the cylinder $\R \times S^1$ is replaced
with an arbitrary Riemann surface $C$.  So we consider an $\CN = 2$
supersymmetric gauge theory and formulate it on $S^2 \times C$.

For a general choice of $C$ supersymmetry is completely broken; the
parameters of the supersymmetry transformation are covariantly
constant spinors (or generalizations thereof), but $C$ admits no such
spinors in general.  In order to preserve some supersymmetry, we must
topologically twist the theory along $C$.  We can do this using a
maximal torus $\U(1)_R$ of the R-symmetry group $\SU(2)_R$.

On $S^2 \times C$, the structure group of the spin connection reduces
to $\U(1)_{S^2} \times \U(1)_C$.  Under $\U(1)_{S^2} \times \U(1)_C
\times \U(1)_R$, the supercharges transform as
\begin{equation}
  (\pm 1, \pm 1, \pm 1).
\end{equation}
The problem is that they have charge $\pm 1$ under $\U(1)_C$, so we
replace $\U(1)_C$ by the diagonal subgroup $\U(1)_C'$ of $\U(1)_C
\times \U(1)_R$.  Then the transformation properties of the
supercharges become
\begin{equation}
  (\pm 1, 0, \pm1) \oplus (\pm 1, 2, 1) \oplus (\pm 1, -2, -1),
\end{equation}
showing that four of them are now scalars on $C$.  The corresponding
supersymmetries now have a chance to survive, since their parameters
can be chosen to be constants on $C$.

It turns out that all of the four supersymmetries do survive on the
curved manifold $S^2$, thanks to the symmetric nature of its geometry.
On the $S^2$, two of the four supercharges are spinors of positive
chirality and the other two are of negative chirality.  Thus we get
$\CN = (2,2)$ supersymmetry on $S^2$ \cite{Benini:2012ui,
  Doroud:2012xw} after the twisting.  The associated transformation
parameters are not covariantly constant spinors on the $S^2$.  Rather,
they are conformal Killing spinors $\veps$, $\vepsb$, obeying the
equations
\begin{equation}
  \label{CKS}
    \nabla_\mu\veps = +\frac{1}{2r}\gamma_\mu\gamma_{\hat3}\veps, \qquad
    \nabla_\mu\vepsb = -\frac{1}{2r}\gamma_\mu\gamma_{\hat3}\vepsb,
\end{equation}
where $\mu = 1$, $2$ is the coordinate index for the $S^2$.%
\footnote{Our conventions for spinors on $S^2$ are as follows.  We use
  spherical coordinates $(x^1, x^2) = (\theta, \varphi)$ on $S^2$ such
  that the round metric of radius $r$ is $r^2\rmd\theta^2 +
  r^2\sin^2\!\theta \, \rmd\varphi^2$.  The hatted index $\muh =
  \hat1$, $\hat2$ refers to the orthonormal frame $e_{\hat1} =
  \del_1/r$, \ $e_{\hat2} = \del_2/r\sin\theta$.  Often we extend
  $\muh$ to run from $\hat1$ to $\hat3$.  The gamma matrices
  $\gamma_{\muh}$ are given by the Pauli matrices, and the chirality
  operator is $\gamma_{\hat3}$.  The product of Dirac spinors
  $\psi\chi = \psi^T C \chi$, with $C = i\gamma_{\hat2}$.  The spin
  connection is denoted by $\nabla$.}
Each of these equations has two independent solutions, so in total we
have four, $\veps_\alpha$, $\vepsb_\alpha$, $\alpha = 1$, $2$.  We
write $\Qb_\alpha$, $Q_\alpha$ for the supercharges corresponding to
$\veps_\alpha$, $\vepsb_\alpha$, and $\CQb_\alpha$, $\CQ_\alpha$ for
their action on fields, respectively.

In addition to the four supersymmetries generated by $\Qb_\alpha$,
$Q_\alpha$, the $\CN = (2,2)$ supersymmetry group contains the
rotations of the $S^2$, and also a $\U(1)$ R-symmetry, which we choose
to be the vector R-symmetry $\U(1)_V$.  (So we are considering A-type
supersymmetry~\cite{Hori:2013ika}.)  The R-symmetry rotates
$\Qb_\alpha$ by charge $q = +1$ and $Q_\alpha$ by $q = -1$.  The
nonvanishing commutators among the supercharges are
\begin{equation}
  \{\Qb_\alpha, Q_\beta\}
  = \CL_\xi + i\alpha F_V
\end{equation}
modulo gauge transformations.  On the right-hand side appear the Lie
derivative $\CL_\xi$ by the Killing vector field $\xi^\mu =
i\veps_\alpha\gamma^\mu\vepsb_\beta$, as well as the $\U(1)_V$
generator $F_V$ accompanied with the parameter $\alpha =
\veps_\alpha\gamma_{\hat3}\vepsb_\beta/2r$.  Note that the commutators
cannot generate translations along $C$, since our supercharges are
scalars on $C$.  As a result, the commutation relations remain
unchanged from the two-dimensional case, even though we are really
dealing with a four-dimensional theory on $S^2 \times C$.

We would like to repackage the field content of the twisted theory
into supermultiplets of $\CN = (2,2)$ supersymmetry.  In general
$\U(1)_R$ is the only $\U(1)$ R-symmetry present in the twisted
theory, so this is identified with $\U(1)_V$.  (There is another
$\U(1)$ R-symmetry if the theory is superconformal.)  The fact that
the vector multiplet scalar is neutral under $\U(1)_R$ means that the
theory should be formulated using vector and chiral multiplets, as
opposed to twisted vector and twisted chiral multiplets.  Unlike the
case of flat spacetime, twisted and untwisted multiplets are
inequivalent representations on $S^2$.

The supersymmetry transformation rules and the supersymmetric action
for the twisted theory can be obtained by lifting the relevant
formulas from two dimensions.  This is relatively straightforward, and
carried out in appendix \ref{N=2-SGT}.

Although the details of this construction will not be needed for our
discussion, one point deserves to be mentioned.  The twisted theory
has four supercharges, and any of their linear combinations can be
used as a BRST operator.  However, for particular linear combinations,
the theory becomes independent of the K\"ahler structure on $C$.  If
we choose the parameters in such a way that $\vepsb_\alpha =
\gamma_{\hat 3} \veps_\alpha$ and $\veps_1\veps_2 = -\vepsb_1\vepsb_2
= 1$, then the relevant linear combinations are $\Qb_1 + \zeta Q_2$
and $Q_1 + \zeta\Qb_2$ with $\zeta \in \C^\times$.  For definiteness
we set
\begin{equation}
  Q = \Qb_1 + Q_2
\end{equation}
and use this as a BRST operator.  This squares to a rotation of the
$S^2$ about the axis through the poles $\theta = 0$ and $\pi$, plus a
vector R-rotation:
\begin{equation}
  Q^2 = \frac{1}{r} \Bigl(\CL_{\del_\varphi} + \frac{1}{2} F_V\Bigr).
\end{equation}
Near the north pole $\theta = 0$, the action of $Q$ looks like that of
a supercharge in the $\Omega$-deformed, topologically twisted theory
\cite{Nekrasov:2002qd} on $\R^2_\veps \times \R \times S^1$ with
$\veps = 1/r$.  Near the south pole $\theta = \pi$, it looks like the
action of the corresponding supercharge in the $\Omega$-deformed
theory with $\veps = -1/r$, twisted in the opposite manner.

Since the $Q$-invariant sector of the twisted theory is invariant
under deformations of the K\"ahler structure of $C$, we can rescale
the metric of $C$ by a large factor.  Then the theory at energies $\mu
\ll 1/r$ is described by a two-dimensional theory on $C$ which depends
only on the conformal structure (for a given spin structure).  The
compactification of this two-dimensional conformal field theory on a
circle is to be identified with the quantum integrable system which we
are after.

\subsection{Infrared theory}

Let us specialize to the case where $C$ is a cylinder $\R \times S^1$,
and consider the low-energy dynamics of the theory.  We take the radii
$r$ of the $S^2$ and $R$ of the $S^1$ to be sufficiently large; in
particular, $r$, $R \gg 1/\Lambda$.  We also take $r \gg R$.  Then, at
energies $\mu \ll \Lambda$ but $\mu \gg 1/R$, the effects of $r$ and
$R$ being finite are small, so the system is described by an effective
abelian theory on $S^2 \times \R \times S^1$ as in the case of flat
spacetime $\R^4$ or $\R^3 \times S^1$.  Its prepotential $\CF$ may
depend on $r$ and $R$, among other parameters of the ultraviolet
theory, and coincides with the prepotential for $\R^4$ in the limit
$r$, $R \to \infty$.

If we further lower the energy scale so that $1/r \ll \mu \ll 1/R$,
then the dynamics can be described by a three-dimensional gauge theory
on $S^2 \times \R$ which, roughly speaking, is the dimensional
reduction of the four-dimensional theory on the $S^1$.  As in the case
of flat spacetime, we dualize the gauge fields in this theory to
periodic scalars.  This step works just as before (since $S^2 \times
\R$ is topologically almost $\R^3$, only the origin removed), and
produces an $\CN = 4$ supersymmetric sigma model whose target space is
the total space of the complex integrable system $\CM \to \CB$.

This sigma model has $\CN = (2,2)$ supersymmetry on $S^2$, as the
ultraviolet theory has this symmetry.  Before the dualization, the
vector multiplet scalars $a^I$ sit in gauge-invariant twisted chiral
multiplets, commonly denoted as $\Sigma^I$ \cite{Hori:2003ic}.  After
the dualization they are again part of twisted chiral multiplets, and
moreover, the same is true for the holomorphic coordinates $z_I$ of
the fibers of $\CM$.  The reason is that, as we will see, in order to
formulate the sigma model we need to turn on a (twisted)
superpotential.  The scalars $a^I$, $z_I$ have vector R-charge $q =
0$, so any superpotential constructed out of them has $q = 0$.  It
follows that if they were part of untwisted chiral multiplets, then
the superpotential would break $\U(1)_V$ and hence supersymmetry.  (A
superpotential breaks $\U(1)_V$ unless it has $q = 2$.  By contrast, a
twisted superpotential preserves $\U(1)_V$ regardless of the vector
R-charge.)

In summary, the low-energy dynamics of the theory on $S^2 \times \R
\times S^1$ is described by an $\CN = 4$ supersymmetric sigma model
with hyperk\"ahler target space $\CM$, formulated on $S^2 \times \R$.
It preserves $\CN = (2,2)$ supersymmetry on the $S^2$ and is
constructed from twisted chiral multiplets.  Our next task is to write
down the action of this sigma model.

\subsection{The sigma model}

The strategy for determining the action of the infrared sigma model on
$S^2 \times \R$ is basically the same as the one we employed for the
ultraviolet theory.  First we write down the action for the
two-dimensional theory on $S^2$ obtained by dimensional reduction on
the $\R$.  Then we lift it to $S^2 \times \R$.

The dimensional reduction gives an $\CN = (2,2)$ supersymmetric sigma
model on $S^2$, with target space $\CM$.  Given holomorphic
coordinates on $\CM$, the map $\ups\colon S^2 \to \CM$ of the sigma
model can be described locally by complex scalars $\ups^i$, $i = 1$,
$\dotsc$, $2r$.  A choice of a local symplectic basis $\{\alpha_I,
\beta^I\}$ of $\Gamma$ provides holomorphic coordinates in the complex
structure $J_3$, namely $(a^I, z_I)$.  So let us focus on this complex
structure.

The scalars $\ups^i$ are completed with Weyl spinors $\chi_+^\ib$,
$\chi_-^i$, $\chib_+^i$, $\chib_-^\ib$ and complex auxiliary fields
$E^i$ to form twisted chiral multiplets; the subscripts $\pm$ of the
spinors indicate the chirality.  Their supersymmetry transformations
are \cite{Gomis:2012wy, Jia:2013foa}
\begin{equation}
  \label{TCM-SUSY}
  \begin{aligned}
    \delta\ups^i &= \vepsb_+\chi_-^i + \veps_-\chib_+^i,
    \\
    \delta\upsb^\ib &= -\vepsb_-\chi_+^\ib - \veps_+\chib_-^\ib,
    \\
    \delta\chi_+^\ib &= i\nablas_+{}^-\upsb^\ib\veps_- - \Eb^\ib\veps_+,
    \\
    \delta\chi_-^i &= i\nablas_-{}^+\ups^i\veps_+ - E^i\veps_-,
    \\
    \delta\chib_+^i &= -i\nablas_+{}^-\ups^i\vepsb_- + E^i\vepsb_+,
    \\
    \delta\chib_-^\ib &= -i\nablas_-{}^+\upsb^\ib\vepsb_+ + \Eb^\ib\vepsb_-,
    \\
    \delta E^i &= i\vepsb_-\nablas_+{}^-\chi_-^i + i\veps_+\nablas_-{}^+\chib_+^i,
    \\
    \delta \Eb^\ib &= -i\vepsb_+\nablas_-{}^+\chi_+^\ib - i\veps_-\nablas_+{}^-\chib_-^\ib.
  \end{aligned}
\end{equation}
Here $\nablas_+{}^-$, $\nablas_-{}^+$ are the nonzero matrix elements
of the Dirac operator $\nablas$.  Note that we are taking $\veps$,
$\vepsb$ to be commuting spinors.

The Lagrangian for the two-dimensional sigma model can be written
compactly in terms of a K\"ahler potential $K$, which is a
locally-defined function on $\CM$ that gives the K\"ahler form
$\omega_3 = ig_{i\jb} \, \rmd\ups^i \wedge \rmd\upsb^\jb$ by $\omega_3
= i\del\delb K$:
\begin{equation}
  \label{L_Ct}
  \begin{split}
    \CL_{\Ct}
    &= \frac12 \bigl(\CQb_1\CQb_2\CQ_1\CQ_2
                           + \CQ_1\CQ_2\CQb_1\CQb_2\bigr)K \\
    &= \frac12 \CQb_1\CQb_2\bigl(g_{i\jb}\chi_-^i\chi_+^\jb\bigr)
       + \frac12\CQ_1\CQ_2\bigl(g_{i\jb}\chib_+^i\chib_-^\jb\bigr).
  \end{split}
\end{equation}
Computing the supersymmetry variations and integrating out the
auxiliary fields, we get
\begin{equation}
  \label{L_Ct-2}
  \CL_{\Ct}
  = g_{i\jb} \del^\mu\ups^i \del_\mu\upsb^\jb
    - ig_{i\jb} \Ds_-{}^+\chib_+^i \chi_+^\jb
    - ig_{i\jb} \chi_-^i \Ds_+{}^-\chib_-^\jb
    + R_{i\jb k\lb} \chi_-^i\chi_+^\jb\chib_+^k\chib_-^\lb.
\end{equation}
The Dirac operator $\Ds$ is coupled to the pullback of the metric
connection of $\CM$ by $\ups$.

To lift the supersymmetry transformations to three dimensions, we just
need to allow the fields to vary along the extra $x^3$-direction; thus
the form of the transformation rules remains unchanged from the
formula \eqref{TCM-SUSY}.

To lift the action, in addition we integrate the two-dimensional
action over the $x^3$-direction:
\begin{equation}
  \label{S_Ct}
  S_{\Ct} = \int_\R \vol_\R \int_{S^2} \vol_{S^2} \, \CL_{\Ct}.
\end{equation}
The symbol $\vol_M$ denotes the volume form of a Riemannian manifold
$M$, and again, the form of $\CL_{\Ct}$ remains unchanged from the
formula \eqref{L_Ct} or \eqref{L_Ct-2}.  However, some terms are
missing from the action $S_{\Ct}$ so obtained, such as kinetic terms
involving derivatives along the $x^3$-direction.  These missing terms
need to be supplied by a twisted superpotential.

In our context, a twisted superpotential is a holomorphic functional
$\Wt$ on $\Map(\R, \CM)$, the space of maps from $\R$ to $\CM$.  The
bosonic field $\ups\colon S^2 \times \R \to \CM$ of the
three-dimensional sigma model gives rise to a map $\upst\colon S^2 \to
\Map(\R, \CM)$, and $\Wt$ is to be understood as a functional of
$\upst$.  Then the twisted F-term is given by
\begin{equation}
  \CL_{\Wt}
  = i\Bigl(
    E \intprod \delta\Wt
    + \chi_- \intprod \delta(\chib_+ \intprod\delta \Wt)
    + \Eb \intprod \delta\Wt^*
    + \chib_- \intprod \delta(\chi_+ \intprod \delta\Wt^*)
    + \frac{2}{r} \Im \Wt\Bigr),
\end{equation}
where $\delta$ is the exterior functional derivative and $\intprod$ is
the interior product.  (For example, $E \intprod \delta\Wt$ means
taking the variation of $\Wt$ under $\upst \to \upst + \delta\upst$,
followed by substitution $\delta\upst = E$.)  The three-dimensional
action therefore takes the form
\begin{equation}
  S = S_{\Ct} + \int_{S^2} \vol_{S^2} \, \CL_{\Wt}.
\end{equation}
With the twisted superpotential $\Wt$ turned on, integrating out the
auxiliary fields produces the potential term $\|\delta\Wt\|^2$.  We
want to choose $\Wt$ in such a way that this potential provides the
bosonic kinetic term involving $x^3$-derivatives.

We expect $\Wt$ to be constructed from the holomorphic symplectic form
$\Omega$, since this is the only object associated with the
hyperk\"ahler structure of $\CM$ that is holomorphic in $J_3$ and can
be integrated in some manner to define a functional.  The appropriate
choice turns out to be the following.  Suppose we have a functional
$\CA(\upst)$ such that under the variation $\upst \to \upst +
\delta\upst$, it changes by
\begin{equation}
  \label{deltaA}
  \delta\CA(\upst) = \int_{\R} \Omega_{ij} \delta\upst^i \rmd\upst^j.
\end{equation}
Given such a functional, we set
\begin{equation}
  \label{Wt}
  \Wt = \frac{i}{2} \CA.
\end{equation}
With this choice, the potential
\begin{equation}
  \|\delta\Wt\|^2
  = \frac14 \int_\R \vol_\R \,
    g^{i\jb} \bigl(\Omega_{ik} \del^3\ups^k\bigr)
            \bigl(\Omegab_{\jb\lb} \del_3\upsb^\lb\bigr).
\end{equation}
Using the relations $\Omega = -i(\omega_1 + i\omega_2)$ and
$\omega_\alpha = J_\alpha g$, we can rewrite this as
\begin{equation}
  \frac14 \int_\R \vol_\R \,
  g\bigl((J_1 + iJ_2) \del^3\ups, \overline{(J_1 + iJ_2) \del_3\ups}\bigr)
  = \frac12 \int_\R \vol_\R \,
    g\bigl(\del^3\ups, \del_3\upsb\bigr).
\end{equation}
In this equality we used the fact that the hermitian metric is
compatible with the complex structure $(J_1 + iJ_2)/\sqrt{2}$.  We see
that this is precisely the missing bosonic kinetic term.  So this is
the right choice for $\Wt$, up to an overall phase.  It will become
clear shortly that the phase is also right.

We now have to construct a functional $\CA$ that has the required
property \eqref{deltaA}.  Let us first assume that the cohomology
class $[\Omega] = 0$ so that there exists a one-form $\lambda$ such
that $\Omega = \rmd\lambda$.  This is the case when the hypermultiplet
masses are zero in the ultraviolet.  Then
\begin{equation}
  \CA(\upst) = \int_\R \upst^*\lambda
\end{equation}
possesses the desired property.

When $[\Omega] \neq 0$, the construction is a bit more involved and
proceeds in three steps.  First, we pick a representative
$\upst_0([\upst])$ in each homotopy class $[\upst]$, which is a class
of maps in $\Map(\R, \CM)$ that coincide with $\upst$ at $x^3 =
\pm\infty$ and can be continuously deformed to $\upst$.  Next, given
$\upst \in \Map(\R, \CM)$, we choose a homotopy $\Yt\colon [0,1] \times
\R \to \CM$ between $\Yt_0 = \upst_0([\upst])$ and $\Yt_1 = \upst$.
Finally, we set
\begin{equation}
  \CA = \int_{[0,1] \times \R} \Yt^*\Omega.
\end{equation}
To verify that this definition satisfies the condition \eqref{deltaA},
we can assume that $\delta\upst$ is supported in a sufficiently small
neighborhood in $\CM$ so that we can use a local expression $\Omega =
\rmd\lambda$ to compute the variation.  Then we indeed get
\begin{equation}
  \delta\CA
  = \int_{[0,1] \times \R}  \delta\bigl(\Yt^* \rmd\lambda\bigr)
  = \int_\R \delta(\upst^*\lambda)
  = \int_{\R} \Omega_{ij} \delta\upst^i \rmd\upst^j.
\end{equation}
If we compactify the $\R$ to a circle and consider the contractible
loops, $\CA$ reduces to the symplectic action functional $\CA_H$ for
Hamiltonian $H = 0$, which plays a fundamental role in Floer homology.

The functional $\CA$ is actually not single-valued, as it depends on a
choice of the homotopy $\Yt$.  If we pick another homotopy $\Yt'$, then
$\Delta \Yt = \Yt' - \Yt$ is a map from $S^1 \times \R$ to $\CM$, and $\CA$
changes by
\begin{equation}
  \Delta\CA = \int_{S^1 \times \R} \Delta \Yt^*\Omega.
\end{equation}
Since $\CL_{\Wt}$ contains the term $2i\Im\Wt/r = i\Re\CA/r$, for the path
integral to be well-defined the integral of $i\Re\Delta\CA/r$ over
the $S^2$ must be an integer multiple of $2\pi i$.  The boundary
conditions at infinity effectively collapse the two ends of the
cylinder $S^1 \times \R$ to points, making a two-cycle.  So this
condition is satisfied if
\begin{equation}
  2r [\Re\Omega] \in H^2(\CM;\Z).
\end{equation}
This can be viewed as the condition on the symplectic form in
geometric quantization of the real symplectic manifold $(\CM,
\Re\Omega/\hbar)$, with $\hbar = 1/2r$.  In our context, it means that
the real part of the hypermultiplet masses must be quantized to
integers in the unit of $\hbar$.

Even though the problem of multi-valuedness is resolved, there are
still ambiguities in the definition of $\CA$.  There are two related
ambiguities here.  One is associated with the choice of the
representative paths $\upst_0$.  The other is the values
$\CA(\upst_0)$, which we can set freely since shifting them by
constants does not affect the variation $\delta\CA$.  To fix these
ambiguities we look at how the term $i\Re\CA/r$ arises via the
dualization, in the semiflat approximation.

In the dualization process, we added to the action of the
effective gauge theory the term
\begin{equation}
  -\frac{i}{2\pi} \int_{S^2 \times \R} F^{(3),I} \wedge \rmd\theta_{m,I}
  = -\frac{i}{2\pi} \int_{S^2 \times \R} \vol_{S^2 \times \R} \,
     F^{(3),I}_{\hat1\hat2} \del_{\hat3}\theta_{m,I} + \dotsb.
\end{equation}
We abbreviated terms involving the components of $F^{(3), I}$ other
than $F^{(3), I}_{\hat1\hat2}$.  On the other hand, comparing the
formulas \eqref{L3} for flat spacetime and \eqref{LVM} for flat target
space, we deduce that the Lagrangian contained
\begin{equation}
  \label{L-curved}
  \frac{R}{2} \Im \tau_{IJ}
  \biggl(F_{\hat1\hat2}^{(3),I} + \frac{\Re a^I}{r}\biggr)
  \biggl(F_{\hat1\hat2}^{(3),J} + \frac{\Re a^J}{r}\biggr)
  + \frac{i}{2\pi} \biggl(\Re\tau_{IJ} F^{(3),I}_{\hat1\hat2}
    + \Im\tau_{IJ} \frac{\Im a^I}{r}\biggr)\del_{\hat3}\theta_e^J.
\end{equation}
Integrating $F_{\hat1\hat2}^I$ out then produces the term
\begin{multline}
  \frac{i}{2\pi r} \bigl[
  \Re a^I \bigl(\del_{\hat3}\theta_{m,I}
                - \Re\tau_{IJ}\del_{\hat3}\theta_e^J\bigr)
  + \Im a^I \Im\tau_{IJ} \del_{\hat3}\theta_e^J\bigr] \\
  = \frac{i}{2\pi r} \Re\bigl[a^I (\del_{\hat3}\theta_{m,I}
                - \tau_{IJ}\del_{\hat3}\theta_e^J\bigr)\bigr].
\end{multline}
This is to be identified with $i\Re\CA/r$ (apart from a term involving
$\del_3\tau_{IJ}$ which we have ignored in this analysis).  Recalling
the definition \eqref{z_I} of the holomorphic coordinates $z_I$, we
see that $\CA$ can be written, locally on $\CM$, as
\begin{equation}
  \label{A}
  \CA =  \frac{1}{2\pi} \int_\R a^I \rmd z_I.
\end{equation}
This formula satisfies the condition \eqref{deltaA}, in view of the
local expression \eqref{Omega} of $\Omega$.%
\footnote{Recall that originally the formula \eqref{Omega} for
  $\Omega$ was obtained in the semiflat approximation, and then we
  went on to argue that there are no instanton corrections.  We can
  now make the same statement more precisely as the nonrenormalization
  of $\Wt$.}

\pagebreak

The formula \eqref{A} fixes the aforementioned ambiguities.  For the
choice of representatives $\upst_0$, we can choose each of them to be
a composition of ``horizontal'' paths along which $\rmd z_I = 0$, and
``vertical'' paths along the fibers above fixed points on $\CB$.  The
value of $\CA(\upst_0)$ is equal to the sum of the values assigned to
these component paths.  For horizontal paths, $\CA = 0$, and for
vertical paths, $\CA$ is given by a linear combination of $a^I$
specified by the above formula.

\subsection{Localization}

We are finally ready to localize the path integral for the low-energy
effective theory.  The essential feature of the infrared sigma model
that allows the localization is that the relevant part \eqref{S_Ct} of
the action is $Q$-exact.  Indeed, up to total derivatives we can write
the twisted chiral multiplet Lagrangian \eqref{L_Ct} as
\begin{equation}
  \CL_{\Ct} = \frac12 \CQ
               \bigl[\CQb_2\bigl(g_{i\jb}\chi_-^i\chi_+^\jb\bigr)
               - \CQ_1\bigl(g_{i\jb}\chib_+^i\chib_-^\jb\bigr)\bigr],
\end{equation}
where we used the fact that $\{Q_\alpha, \Qb_\alpha\}$ generates a
rotation of the $S^2$, and the K\"ahler property of the target space
metric.

The $Q$-exactness of $S_{\Ct}$ means that we can freely rescale it by
an overall factor without affecting the $Q$-invariant sector of the
theory.  In particular, we can rescale it as $S_{\Ct} \to t^2 S_{\Ct}$
and take the limit $t \to \infty$.  Then, integrating out the
auxiliary fields leaves no potential term, and the integration over
$\ups$ receives contributions only from a neighborhood of the
configurations such that
\begin{equation}
    \del_\mu\ups^i = 0.
\end{equation}
The path integral therefore localizes to the maps $\ups_0\colon S^2
\times \R \to \CM$ that are constant on the $S^2$.

To evaluate the path integral, we split $\ups$ as $\ups = \ups_0 +
\ups'$, and first integrate over the fluctuations $\ups'$ as well as
the fermions.  (More precisely, $\ups'$ are sections of the pullback
of the tangent bundle of $\CM$ by $\ups_0$.)  At each point on the
$\R$, the integration variables are the modes of the relevant
differential operators.  For $\ups'$, we only integrate over the
nonzero modes since the zero modes just shift the background $\ups_0$
to another one.  As for the fermions, $\CM$ being hyperk\"ahler,
$c_1(\CM) = 0$ and the index of the relevant Dirac operator vanishes.
So there are no fermion zero modes generically.

We can rescale $\ups'$ and the fermions by a factor of $1/t$ so that
the overall factor $t^2$ disappears from the kinetic terms.  After
doing so, the only terms in the action that involve these fields and
survive in the limit $t \to \infty$ are the quadratic terms of
$\CL_{\Ct}$.  For each background $\ups_0$ and at each point on the
$\R$, we can find K\"ahler normal coordinates such that
$g_{i\jb}(\ups_0) = \delta_{ij}$ and $\del_k g_{i\jb}(\ups_0) =
\del_\kb g_{i\jb}(\ups_0) = 0$.  In these coordinates the relevant
part of the Lagrangian is
\begin{equation}
  \sum_i \bigl(\del^\mu\ups'^i \del_\mu\upsb'^\ib
  - i\dels_-{}^+ \chib_+^i \chi_+^\ib
  - i\chi_-^i \dels_+{}^- \chib_-^\ib\bigr).
\end{equation}
Since they are independent of $\ups_0$, the path integral over $\ups'$
and the fermions just produces a constant, which we absorb in the
measure.

The final step in the path integral is to integrate over all possible
backgrounds $\ups_0$.  As these are constant on the $S^2$, the
integration over the $S^2$ just gives a factor of $4\pi r^2$.  Then,
viewing $\ups_0$ as maps from $\R$ to $\CM$, in the end we arrive at
the following path integral of a quantum mechanical system:
\begin{equation}
  \label{QIS-PI}
  \int \cD\ups_0 \exp\Bigl(\frac{i}{\hbar} S(\ups_0)\Bigr).
\end{equation}
Here the action and the Planck constant are given by
\begin{equation}
  S = -2\pi \Re \CA, \qquad
  \hbar = \frac{1}{2r}.
\end{equation}
Locally on $\CM$, the action is expressed as
\begin{equation}
  S = -\int_\R \Re\bigl(a^I \rmd z_I\bigr)
   = -\int_\R \bigl(\Re a^I \rmd\theta_{m,I}
                    - \Re a_{D,I} \, \rmd\theta_e^I\bigr),
\end{equation}
where we used the boundary conditions $\rmd a^I = 0$ at infinity to
obtain the last expression.

The above action is the one for the real integrable system $(\CM,
\Re\Omega)$, written in action-angle variables; there are $2r$
commuting action variables $\Re a^I$, $\Re a_{D,I}$, and $2r$ commuting
angle variables $\theta_{m,I}$, $\theta_e^I$.  We have shown that the
path integral of the $Q$-invariant sector of the effective theory
reduces to the path integral quantizing this classical integrable
system.  Therefore, the low-energy dynamics of the $Q$-invariant
sector is described by the corresponding quantum integrable system.

Let us check semiclasically that the quantum integrable system
reproduces the vacuum structure of the theory on $S^2 \times \R \times
S^1$.  Suppose that we fix the holonomies $\theta_e^I$ at infinity.
Then the effect of the curvature to the vacuum moduli is that $a^I$
must satisfy
\begin{equation}
  \Re a^I \in \frac{\Z}{2r}.
\end{equation}
This is due to flux quantization and the fact that the gauge kinetic
term $\Tr F_{\hat1\hat2}^2$ is shifted to $\Tr(F_{\hat1\hat2} +
\Re\phi/r)^2$ in the ultraviolet Lagrangian \eqref{L-S2xC}.  This
condition is recovered in the quantum integrable system from the
constraint
\begin{equation}
  \frac{\Re a^I}{\hbar} \in \Z
\end{equation}
obtained by integrating over the periodic scalars $\theta_{m,I}$.  If
we instead chose to fix $\theta_{m,I}$ and integrate over
$\theta_e^I$, then we would get the electromagnetic dual of the above
constraint.

\section{The hemisphere case}
\label{hemisphere}

Lastly, let us discuss what happens when the sphere $S^2$ in the
spacetime is replaced with a hemisphere $D^2$ of radius $r$.  Recall
that the square of our supercharge $Q = \Qb_1 + Q_2$ generates a
rotation of the $S^2$.  We take $D^2$ to be invariant under this
rotation.

The supersymmetry transformations and the supersymmetric Lagrangian
are the same as in the $S^2$ case.  The new feature is that the
spacetime has a boundary, so we have to specify a boundary condition
that preserves $Q$.  We also demand that it preserves the rotational
symmetry of $D^2$.  As $\Qb_1$ and $Q_2$ have opposite charges under
the rotation, such boundary conditions preserve these supercharges
separately.  Thus they are half-BPS boundary conditions of the $\CN =
(2,2)$ supersymmetry, describing half-BPS branes in the target space.
$\CN = (2,2)$ supersymmetric gauge theories on a hemisphere with
half-BPS boundary conditions have recently been studied in
\cite{Sugishita:2013jca, Honda:2013uca, Hori:2013ika}.

Of particular interest to us are branes supported on the
middle-dimensional submanifolds $\CL_1$, $\CL_2 \subset \CM$ defined
by
\begin{align}
  \CL_1&\colon \Im a_{D,I} = 0 = \theta_{m,I}, \\
  \CL_2&\colon \Im a^I = 0 = \theta_e^I.
\end{align}
Since $\Omega \propto \rmd a^I \wedge \rmd\theta_{m,I} - \rmd a_{D,I}
\wedge \rmd\theta_e^I$, these submanifolds are Lagrangian with respect
to $\omega_1 = -\Im\Omega$.  In the semiflat approximation one can
check that they are holomorphic under $J_2$ and Lagrangian with
respect to $\omega_3$.  The same kinds of branes were studied by
Nekrasov and Witten \cite{Nekrasov:2010ka} to establish a connection
between $\CN = 2$ supersymmetric gauge theories on the
$\Omega$-deformed spacetime $\R^2_\veps \times \R \times S^1$ and
quantum integrable systems.  There is a similar connection in the
present setup.

Just as in the $S^2$ case, we can show that the $Q$-invariant sector
of the low-energy effective theory on $D^2 \times \R \times S^1$ is
described by a quantum integrable system.  The path integral localizes
to the configurations $\ups_0$ that are constant on $D^2$ and
therefore determined by the boundary value.  These are maps from $\R$
to $\CL \subset \CM$, where $\CL = \CL_1$ or $\CL_2$ depending on the
choice of the boundary condition.  The one-loop determinants are still
independent of the background configuration $\ups_0$ and can be
absorbed in the measure.  The value of the action for $\ups_0$ is half
of that in the $S^2$ case, since the area of the spacetime is half.
Hence, the localization leads to the same expression \eqref{QIS-PI},
with the differences being that the integration domain is now
$\Map(\R, \CL)$ and the Planck constant is twice the previous value:
\begin{equation}
  \hbar = \frac{1}{r}.
\end{equation}
We conclude that the result of the localization is the path integral
for a quantum integrable system that quantizes the real integrable
system $(\CL, \Re\Omega)$.

The Hilbert space of the quantum integrable system is associated to a
``time slice'' at fixed $x^3$.  So physical states are described in
the gauge theory as $Q$-invariant functionals of field configurations
over $D^2 \times \{x^3\} \times S^1$.  We can recast these states to
states of open strings stretched between two branes.  For this, we
reduce the theory on the circle fibers of $D^2$, in addition to the
reduction on the $S^1$ which we have been considering.  This
additional reduction turns $D^2$ into an interval $I = [0, r]$, and
the theory becomes a sigma model on $I \times \R$.  We now have two
branes, located at the two ends of $I$.  One of them is the brane we
placed on the boundary of $D^2$.  The other, new brane sits at the end
that was formerly the pole of $D^2$.  This is a space-filling brane
since the pole was not constrained to be mapped to any submanifold of
$\CM$.  In this process of reduction, the gauge theory states are
turned into open string states stretched between these two branes.  We
see here a close parallel to the construction of Nekrasov and Witten;
in their construction, one reduces the $\Omega$-deformed theory on the
circle fibers of a cigar-shaped manifold (which looks much like a
hemisphere near the tip) to arrive at a topological sigma model on $\R
\times I$ with target space $\CM$, and the Hilbert space of the
quantum integrable system is obtained as the space of open strings
stretched between a space-filling $(A,B,A)$-brane and a
middle-dimensional $(A,B,A)$-brane located at the ends of $I$.

The effective prepotential determines the spectrum of the quantum
integrable system in the form of the Bethe ansatz equation.  As an
example, take $\CL = \CL_1$.  The action of the quantum integrable
system is then
\begin{equation}
  \label{QI-action}
  S = \int_\R \Re a_{D,I} \, \rmd\theta_e^I.
\end{equation}
Since the $\Re a_{D,I}$ commute with one another, states are labeled
by their eigenvalues.  Integrating over the periodic scalars
$\theta_e^I$ imposes the constraint
\begin{equation}
  \frac{\Re a_{D,I}}{\hbar} \in \Z
\end{equation}
on the possible values of these parameters.  In view of the fact that
$\Im a_{D,I} = 0$ on $\CL$, this condition can be written as
\begin{equation}
  r a_{D,I}
  = r \frac{\del\CF(a; r)}{\del a^I}
  \in \Z,
\end{equation}
This is the Bethe ansatz equation with Yang-Yang function $Y =
r\CF/2\pi i$.

What we have just found is a variant of the correspondence discovered
by Nekrasov and Shatashvili \cite{Nekrasov:2009rc}.  The
$\Omega$-deformed spacetime $\R^2_\veps \times \R \times S^1$ reduces
in the infrared to a two-dimensional gauge theory on $\R \times S^1$.
If we write $\CW(a; \veps)$ for the twisted superpotential of this
theory, then the equation that determines the vacua is
\begin{equation}
  \frac{\del\CW(a; \veps)}{\del a^I} \in i\Z.
\end{equation}
The Nekrasov-Shatashvili correspondence identifies $\CW$ with the
Yang-Yang function of the quantum integrable system.%
\footnote{In their case the correspondence can be established by
  considering a topological field theory, so the states of the quantum
  integrable system have zero energy and correspond to the vacua of
  the gauge theory.  This is not the case for us, even though the
  action \eqref{QI-action} appears to suggest that the Hamitonian is
  zero.  The reason is that in the localization of path integral we
  ignored the ratio of the one-loop determinants, which shifts the
  Lagrangian by a zero-point energy.  The energy becomes zero only in
  the limit $r \to \infty$, where the determinants for scalars and
  spinors are equal.}
We see that $\CW$ plays the role of $r\CF$ in our correspondence.

The two correspondences agree in the limit $r \to \infty$ and $\veps
\to 0$.  In the limit $\veps \to 0$, the twisted superpotential
behaves as
\begin{equation}
  \CW(a; \veps) = \frac{i\CF(a; \veps = 0)}{\veps} + \dotsb,
\end{equation}
where $\CF(a; \veps)$ is the effective prepotential of the
$\Omega$-deformed theory, and $\dotsb$ denotes terms regular in
$\veps$.  Since $\CF(a; \veps = 0)$ is the effective prepotential on
flat spacetime $\R^3 \times S^1$ and therefore equals $\CF(a; r =
\infty)$, their correspondence coincides with ours in this limit under
the identification $\veps = 1/r$.

\section*{Acknowledgments}

We would like to thank Kazuo Hosomichi and Takuya Okuda for helpful
comments.  This work is supported by National University of Singapore
Start-up Grant R144-000-269-133.

\appendix

\section{\texorpdfstring{$\CN = 2$}{N = 2} supersymmetric gauge
  theories on $S^2 \times C$}
\label{N=2-SGT}

In this appendix we formulate $\CN = 2$ supersymmetric gauge theory on
$S^2 \times C$, with $C$ a Riemann surface.  We equip the $S^2$ with a
round metric of radius $r$, and $C$ with a K\"ahler metric $h$.

As explained in section~\ref{UV}, the theory is twisted along $C$ and
possesses $\CN = (2,2)$ supersymmetry on $S^2$.  So we can write down
the supersymmetry transformation rules and supersymmetric Lagrangians
following the general prescription for $\CN = (2,2)$ supersymmetric
gauge theories on $S^2$ \cite{Benini:2012ui, Doroud:2012xw}.  First of
all, we need to understand how the vector multiplet and
hypermultiplets decompose as supermultiplets of $\CN = (2,2)$
supersymmetry.

Let us start with the vector multiplet.  After the twisting, four
components $\lambda$, $\lambdab$ of the gauginos become scalars on $C$
and Dirac spinors on $S^2$.  Together with the vector multiplet scalar
$\phi = \phi_1 + i\phi_2$, the components $A_\mu$, $\mu = 1$, $2$, of
the gauge field along $S^2$, and a real auxiliary field $D$, they form
an $\CN = (2,2)$ vector multiplet $V$:
\begin{equation}
  V = (\phi, \lambda, \lambdab, A_\mu, D).
\end{equation}
The rest of the $\CN = 2$ vector multiplet fields are divided into two
groups according to their transformation properties under $\U(1)_C'$.
We choose a holomorphic coordinate $z$ on $C$ such that $(1,0)$-forms
have charge $-2$.  Then, one group form an $\CN = (2,2)$ chiral
multiplet $\Phi_z$ of R-charge $q = 0$ in the adjoint representation
together with a complex auxiliary field $F_z$, while the other form
the corresponding antichiral multiplet $\Phib_\zb$:
\begin{equation}
    \Phi_z = (A_z, \lambda_z, F_z), \qquad
    \Phib_\zb = (A_\zb, \lambdab_\zb, \Fb_\zb).
\end{equation}
Our convention for chiral multiplets is that if the scalar component
has R-charge $q$, then the spinor has R-charge $q-1$.

Now we turn to hypermultiplets.  A hypermultiplet consists of two $\CN
= 1$ chiral multiplets.  If we write $M$ and $\Mt^\dagger$ for the
scalars of these chiral multiplets and assign them R-charge $q = +1$
and $-1$, then after the twisting they become sections $M_+$ and
$\Mt^\dagger_-$ of $\Kb_C^{1/2}$ and $K_C^{1/2}$, respectively.  These
are part of a chiral multiplet $H_+$ and an antichiral multiplet
$\Ht^\dagger_-$, both in the same representation $R$ which is the
representation of the hypermultiplet:
\begin{equation}
  H_+ = (M_+, \psi_+, F_+), \qquad
  \Ht^\dagger_- = (\Mt^\dagger_-, \psit^\dagger_-, \Ft^\dagger_-).
\end{equation}
Their hermitian conjugates are part of an antichiral multiplet
$H^\dagger_-$ and a chiral multiplet $H_+$ in the dual representation
$R^\vee$:
\begin{equation}
  H^\dagger_- = (M^\dagger_-, \psi^\dagger_-, F^\dagger_-), \qquad
  \Ht_+ = (\Mt_+, \psit_+, \Ft_+).
\end{equation}

The supersymmetry transformation rules for these multiplets are as
follows: for $V$,%
\footnote{Our definition of $D$ differs from that in
  \cite{Benini:2012ui} by the shift $D \to D + \phi_2/r$.}
\begin{equation}
  \begin{aligned}
   \delta A_\mu &= -\frac{i}{2} \bigl(\vepsb\gamma_\mu\lambda
                                + \veps\gamma_\mu\lambdab\bigr), \\
   \delta\phi &= \vepsb\gamma_-\lambda - \veps\gamma_+\lambdab, \\
   \delta\phib &= \vepsb\gamma_+\lambda - \veps\gamma_-\lambdab, \\
   \delta\lambda
   &= i\Bigl[\Bigl(F_{\hat1\hat2} + \frac{\phi_1}{r}\Bigr) \gamma_{\hat3}
            + \gamma_-\Ds\phi
            + \gamma_+\Ds\phib
            + \frac{1}{2} [\phi, \phib] \gamma_{\hat3}
            + iD\Bigr]\veps, \\
   \delta\lambdab
   &= i\Bigl[\Bigl(F_{\hat1\hat2} + \frac{\phi_1}{r}\Bigr) \gamma_{\hat3}
             - \gamma_+\Ds\phi
             - \gamma_-\Ds\phib
            - \frac{1}{2} [\phi, \phib] \gamma_{\hat3}
            - iD\Bigr]\vepsb, \\
   \delta D &= -\frac{i}{2} \vepsb
               \bigl(\Ds\lambda
                + [\phi, \gamma_+\lambda]
                + [\phib, \gamma_-\lambda]\bigr)
               + \frac{i}{2} \veps
                 \bigl(\Ds\lambdab
                  - [\phi, \gamma_-\lambdab]
                  - [\phib, \gamma_+\lambdab]\bigr);
\end{aligned}
\end{equation}
for $\Phi_z$, $\Phib_\zb$,
\begin{equation}
  \begin{aligned}
    \delta A_z &= \vepsb\lambda_z, \\
    \delta A_\zb &= \veps\lambdab_\zb, \\
    \delta\lambda_z
    &= \bigl(i\gamma^\mu F_{\mu z} + D_z\phi\gamma_+
             + D_z\phib\gamma_-\bigr)\veps + F_z\vepsb, \\
    \delta\lambdab_\zb
    &= \bigl(i\gamma^\mu F_{\mu\zb} - D_\zb\phi\gamma_-
             - D_\zb\phib\gamma_+)\vepsb  + \Fb_\zb\veps, \\
    \delta F_z
    &= i\veps\bigl(\Ds\lambda_z - \gamma_-[\phi,\lambda_z]
                   - \gamma_+[\phib,\lambda_z] + iD_z\lambda\bigr), \\
    \delta\Fb_\zb
    &= i\vepsb\bigl(\Ds\lambdab_\zb - \gamma_+[\lambdab_\zb,\phi]
               - \gamma_-[\lambdab_\zb,\phib] + iD_\zb\lambdab\bigr);
  \end{aligned}
\end{equation}
and for $H_+$, $H^\dagger_-$,
\begin{equation}
  \begin{aligned}
    \delta M_+ &= \vepsb\psi_+, \\
    \delta M_-^\dagger &= \veps\psi^\dagger_-, \\
    \delta\psi_+ &= i\Bigl(\Ds M_+ + \phi M_+\gamma_+ + \phib M_+\gamma_-
                    + \frac{1}{2r} M_+\gamma_{\hat3}\Bigr)\veps + F_+\vepsb, \\
    \delta\psi^\dagger_- &= i\Bigl(\Ds M_-^\dagger +  M_-^\dagger\phi\gamma_-
                            + M_-^\dagger\phib\gamma_+
                            - \frac{1}{2r} M_-^\dagger\gamma_{\hat3}\Bigr)\vepsb
                     + F^\dagger_-\veps, \\
   \delta F_+ &=  i\veps\Bigl(\Ds\psi_+ - \gamma_-\phi\psi_+
                              - \gamma_+\phib\psi_+ - \lambda M_+
                              + \frac{1}{2r} \gamma_{\hat3} \psi_+\Bigr), \\
   \delta F^\dagger_- &= i\vepsb\Bigl(\Ds\psi^\dagger_- - \gamma_+\psi^\dagger_-\phi
                               - \gamma_-\psi^\dagger_-\phib + M_-^\dagger\lambdab
                               - \frac{1}{2r} \gamma_{\hat3}\psi^\dagger_-\Bigr).
  \end{aligned}
\end{equation}
The supersymmetry transformations for $\Ht_+$, $\Ht^\dagger_-$ are
obtained from those for $H_+$, $H^\dagger_-$ by replacing the fields
appropriately.  In the above formulas, $\gamma_\pm = (1 \pm
\gamma_{\hat3})/2$ are the projectors to the positive and negative
chirality subspaces, and $\Ds = \gamma^\mu D_\mu$ with $D = \nabla -
iA$ the covariant derivative coupled to the spin connection and the
gauge field.

The standard supersymmetric Lagrangians on $S^2$ for vector and chiral
multiplets lift to the following Lagrangians for $V$ and $\Phi_z$,
$\Phib_\zb$:
\begin{align}
  \begin{split}
  \CL_V
  &= \frac12
     \Tr\biggl[
     \Bigl(F_{\hat1\hat2} + \frac{\phi_1}{r}\Bigr)^2
     + D^\mu\phi D_\mu\phib
     + \frac14 [\phi, \phib]^2
     + D^2
     \\ & \hspace{17.93em}
     + i\lambda \bigl(\Ds\lambda + [\phi, \gamma_+\lambda]
     + [\phib,  \gamma_-\lambda]\bigr)\biggr],
  \end{split} \\
  \begin{split}
  \CL_\Phi
  &= 
  \Tr\biggl[F^{\mu z} F_{\mu z}
  + \frac12\bigl(D^z\phi D_z\phib + D^\zb\phi D_\zb\phib\bigr)
  + \Bigl(D + \frac{\phi_2}{r}\Bigr)F^z{}_z + \Fb^z F_z
  \\ & \hspace{8em}
    - i\lambdab^z \bigl(\Ds\lambda_z - [\phi, \gamma_-\lambda_z]
                      - [\phib, \gamma_+\lambda_z]\bigr)
    + \lambdab^z D_z\lambda
    + D^z\lambdab \lambda_z\biggr].
  \end{split}
\end{align}
The $\CN = 2$ vector multiplet action on $S^2 \times C$ is simply
\begin{equation}
  \label{LVM}
  \frac{1}{e^2} \int_{S^2 \times C}
  \vol_{S^2 \times C} (\CL_V + \CL_\Phi)
  + \frac{i\theta}{8\pi^2} \int_{S^2 \times C} F \wedge F.
\end{equation}
Here $\vol_{S^2 \times C}$ is the volume form of $S^2 \times C$.  We
see that the action contains all the required kinetic terms.  In
particular, the $F^{z\zb} F_{z\zb}$ term arises from integrating out
$D$.

For the hypermultiplet, the Lagrangian for $H_+$, $H^\dagger_-$
obtained from the corresponding chiral multiplet Lagrangian in two
dimensions is
\begin{multline}
  \CL_H =
  \sqrt{h^{z\zb}} \biggl[
  D^\mu M^\dagger_- D_\mu M_+
  + M^\dagger_- \Bigl(\frac12 \{\phi, \phib\}
                   + iD + \frac{1}{4r^2}\Bigr) M_+
  + F^\dagger_- F_+ \\
  - i\psi^\dagger_-\Bigl(\Ds - \phi\gamma_- - \phib\gamma_+
                  + \frac{1}{2r} \gamma_{\hat3}\Bigr)\psi_+
  + i\psi^\dagger_-\lambda M_+ - iM^\dagger_-\lambdab\psi_+\biggr].
\end{multline}
The Lagrangian $\CL_{\Ht}$ for $\Ht_+$, $\Ht^\dagger_-$ is similar.
To get the kinetic terms along $C$, we must turn on a superpotential.
Up to an overall phase, the right choice is
\begin{equation}
  W = \sqrt2 h^{z\zb} \Mt_+ D_zM_+.
\end{equation}
This is part of a chiral multiplet whose auxiliary field
\begin{multline}
    F_W
    = \sqrt2 h^{z\zb} \bigl(
       \Ft_+ D_zM_+
       - D_z\Mt_+ F_+
       - i\Mt_+ F_z M_+ \\
       - \psit_+ D_z\psi_+
       + i\psit_+\lambda_z M_+
       + i\Mt_+\lambda_z\psi_+\bigr).
\end{multline}
The complex conjugate $\Wb$ of $W$ is part of an antichiral
multiplet.  If we write $\Fb_W$ for its auxiliary field,
the F-term is given by
\begin{equation}
  \CL_W = i\bigl(F_W + \Fb_W\bigr).
\end{equation}
The hypermultiplet action is then
\begin{equation}
  \frac{1}{e^2} \int_{S^2 \times C} \vol_{S^2 \times C}
  \bigl(\CL_H + \CL_{\Ht} + \CL_W\bigr).
\end{equation}
As usual, hypermultiplet masses can be introduced by weakly gauging
flavor symmetries and giving vacuum expectation values to the vector
multiplet scalars.

After integrating out the auxiliary fields, the bosonic part of the
total Lagrangian becomes
\begin{multline}
  \label{L-S2xC}
  \frac12 \Tr\biggl[\Bigl(F_{\hat1\hat2} + \frac{\phi_1}{r}\Bigr)^2
     + 2F^{\mu z} F_{\mu z}
     + F^{z\zb} F_{z\zb}
     + D^m\phi D_m\phib
     + \frac14 [\phi, \phib]^2
     + \frac{2}{r} \phi_2 F^z{}_z \biggr] \\
  + \sqrt{h^{z\zb}} \Bigl(
    D^m M^\dagger_- D_m M_+
  + D^m\Mt_+ D_m\Mt^\dagger_-
  - M^\dagger_- R^z{}_z M_+
  - \Mt_+ R^z{}_z \Mt^\dagger_- \\
  + \frac{1}{4r^2} \bigl(M^\dagger_- M_+ + \Mt_+\Mt^\dagger_-\bigr)
  + \frac12 M^\dagger_- \{\phi, \phib\} M_+
  + \frac12 \Mt_+ \{\phi, \phib\} \Mt^\dagger_-\Bigr) \\
  + \frac12 \bigl\|M^\dagger_- T_a M_+ - \Mt_+ T_a \Mt^\dagger_-\bigr\|^2
  + 2\|\Mt_+ T_a M_+\|^2,
\end{multline}
where $m$ runs from $1$ to $4$, $R^z{}_z = [\nabla^z, \nabla_z]$, \
$T_a$ are generators of the gauge symmetry in the representation $R$,
and the norm on the Lie algebra is given by the Killing form.  If we
drop the terms with explicit $r$ dependence, this reproduces precisely
the bosonic Lagrangian for the theory on $\R^4$.  Therefore the above
Lagrangian describes the theory formulated on $S^2 \times C$.

We remark that the Lagrangian \eqref{L-S2xC} contains the mass terms
for the hypermultiplet scalars with mass proportional to $1/r$.  So
they are set to zero in vacua; there is no Higgs branch.

The pieces $\CL_V$, $\CL_H$ and $\CL_{\Ht}$ of the total Lagrangian
can be written in $Q$-exact forms for an appropriate choice of a
supercharge $Q$.  For example, we have
\begin{align}
  \CL_V
  &= \frac12 \CQ
      \bigl[\CQ_2 \Tr(\lambdab\lambdab)
            + \zeta^{-1} \CQb_1 \Tr(\lambda\lambda)\bigr],
\\
  \CL_H
  &= \frac12 \CQ
      \bigl[\CQ_2(F^\dagger_- M_+) + \zeta^{-1}\CQb_1(M^\dagger_- F_+)\bigr],
\end{align}
for any $Q = Q_1 + \zeta\Qb_2$ with $\zeta \in \C^\times$.  The other
pieces $\CL_\Phi$ and $\CL_W$ are not $Q$-exact.  (A formula similar
to the one for $\CL_H$ would not work for $\CL_\Phi$, since the scalar
$A_z$ of $\Phi_z$ is not a globally-defined object.)  Nevertheless,
these terms do not introduce dependence on the K\"ahler structure of
$C$, since the volume form of $C$ is given by $\vol_C = ih_{z\zb} \rmd
z \wedge \rmd\zb$ and $\vol_C \, h^{z\zb}$ is independent of $h$.  It
follows that the twisted theory is independent of the K\"ahler
structure if we regard $Q$ as a BRST operator.

Since hypermultiplets are spinors on $C$ after the twisting,
formulating the twisted theory requires picking a spin structure on
$C$.  We can avoid this by redefinition of the $\U(1)_R$ symmetry used
in the twisting.  The theory has a global symmetry $\U(1)_B$ under
which $H$ and $\Ht$ have opposite charges.  We can shift $\U(1)_R$ by
$\U(1)_B$ so that the hypermultiplets have integer R-charges, say $q =
2$ for $H$ and $q = 0$ for $\Ht$.  Then the twisting turns $H$ into a
$(0,1)$-form and $\Ht$ into a scalar on $C$.  For this vector R-charge
assignment,%
\footnote{Actually there is no fundamental reason that we must equate
  $\U(1)_V$ and $\U(1)_R$, as there can be a shift by a global $\U(1)$
  symmetry.  However, if they are different, the action of $Q$ near
  the poles can no longer be interpreted as the action of a
  supercharge of the twisted $\Omega$-deformed theory.}
there are no mass terms due to the curvature of $S^2$ and there can be
a Higgs branch.

\bibliographystyle{JHEP}
\bibliography{N=2-QIS}
\end{document}